\documentclass{aa}
\usepackage{psfig,graphicx,ulem,epsfig}
\usepackage{txfonts}
\usepackage{morefloats}
\usepackage{soul,color}

\begin{document}

\def\kart{Kartaltepe et al. (2008)}
\def\karp{Kartaltepe et al. 2008}

\title{Searching for filaments and large-scale structure around
  DAFT/FADA clusters \thanks{Based on our own data (see Guennou et
    al. 2014) and archive data obtained with MegaPrime/MegaCam, a
    joint project of CFHT and CEA/DAPNIA, at the Canada-France-Hawaii
    Telescope (CFHT) which is operated by the National Research
    Council (NRC) of Canada, the Institute National des Sciences de
    l'Univers of the Centre National de la Recherche Scientifique of
    France, and the University of Hawaii, and based on
    Subaru/SuprimeCam archive data. Also based on observations made with
    the Gran Telescopio Canarias (GTC), installed in the Spanish
    Observatorio del Roque de los Muchachos of the Instituto de
    Astrof\'\i sica de Canarias, in the island of La Palma.  This
    research has made use of the NASA/IPAC Extragalactic Database
    (NED) which is operated by the Jet Propulsion Laboratory,
    California Institute of Technology, under contract with the
    National Aeronautics and Space Administration.  }}

\author{
F.~Durret \inst{1} \and
I.~M\'arquez \inst{2} \and
A.~Acebr\'on \inst{3} \and
C.~Adami \inst{3} \and
A.~Cabrera-Lavers \inst{4} \and
H.~Capelato \inst{5,6} \and
N.~Martinet \inst{1,7}\and
F.~Sarron \inst{1} \and 
M.P.~Ulmer \inst{8}
}

\institute{
Sorbonne Universit\'es, UPMC Univ. Paris 6 et CNRS, UMR~7095, Institut d'Astrophysique de Paris, 
98bis Bd Arago, 75014, Paris, France
\and
Instituto de Astrof\'isica de Andaluc\'ia, CSIC, Glorieta de la Astronom\'ia s/n, 18008, Granada,
Spain
\and
Aix Marseille Universit\'e, CNRS, LAM (Laboratoire d'Astrophysique de Marseille) 
UMR~7326, 13388, Marseille, France 
%LAM, OAMP, P\^ole de l'Etoile Site Ch\^ateau-Gombert, 38 rue Fr\'ed\'eric 
%Joliot--Curie,  13388 Marseille Cedex 13, France
\and
Instituto de Astrof\'isica de Canarias, V\'ia L\'actea s/n, 38205 
La Laguna, Tenerife, Spain
\and
Divis\~ao de Astrof\'\i sica, INPE/MCT, 12227-010 S\~ao Jos\'e dos Campos, 
S\~ao Paulo, Brazil
\and
N\'ucleo de Astrof\'\i sica Te\'orica, Universidade Cruzeiro do Sul, Rua 
Galv\~ao Bueno, 868, 01506-000 S\~ao Paulo, Brazil
\and
Argelander Institute for Astronomy, University of Bonn, Auf dem
H\"ugel 71, D-53121 Bonn, Germany
\and
Department of Physics $\&$ Astronomy, CIREA, Northwestern University, Evanston, IL 60208-2900, USA 
}

\date{Accepted: January 22, 2016. Received: October 28, 2015. Draft printed: \today}

\authorrunning{Durret et al.}

\titlerunning{Large scale structure around DAFT/FADA clusters}

\abstract
% context heading (optional)
{Clusters of galaxies are located at the intersection of cosmic
  filaments and are still accreting galaxies and groups along these
  preferential directions. However, because of their relatively low
  contrast on the sky, filaments are difficult to detect (unless a
  large amount of spectroscopic data are available), and unambiguous
  detections have been limited until now to relatively low redshifts
  ($z<\sim0.3$). }
% aims heading (mandatory)
{ This project is aimed at searching for extensions and filaments
    around clusters, traced by galaxies selected to be at the cluster
    redshift based on the red sequence. In the $0.4<z<0.9$ redshift
    range of our sample, clusters are believed to be already well
    formed, but still to be accreting material along filaments.}
% methods heading (mandatory)
{ We have searched for extensions and filaments around the thirty
  clusters of the DAFT/FADA survey for which we had deep wide field
  photometric data. For each cluster, based on a colour-magnitude
  diagram, we selected galaxies that were likely to belong to
    the red sequence, and hence to be at the cluster redshift, and
  built density maps. By computing the background for each of these
  maps and drawing $3\sigma$ contours, we estimated the elongations of
  the structures detected in this way. Whenever possible, we
  identified the other structures detected on the density maps with
  clusters listed in NED.}
% results heading (mandatory)
{We find clear elongations in twelve clusters out of thirty, with
  sizes that can reach up to 7.6~Mpc. Eleven other clusters have
  neighbouring structures, but the zones linking them are not detected
  in the density maps at a 3$\sigma$ level. Three clusters show no
  extended structure and no neighbours, and four clusters are of too
  low contrast to be clearly visible on our density maps.}
% conclusions
{The simple method we have applied appears to work well to show the
  existence of filaments and/or extensions around a number of clusters
  in the redshift range $0.4<z<0.9$. We plan to apply it to other
  large cluster samples such as the clusters detected in the CFHTLS
  and SDSS-Stripe~82 surveys in the near future. }


\maketitle

\section{Introduction}

Starting with the pioneering work of de Lapparent et al. (1986),
observations based on spectroscopic redshifts have shown that galaxies
were not distributed homogeneously on the sky, but were concentrated
along filaments and sheets, with large regions almost devoid of
galaxies. Since then, this result has been confirmed by many surveys,
either shallow but covering large regions on the sky (e.g. 6dF, Jones
et al. 2009; SDSS, Doroshkevich et al. 2004), or deep but limited to
much smaller fields (e.g. VVDS, Le~F\`evre et al. 2005). Numerical
simulations of dark matter particles, which were started in the 1970s by Press
\& Schechter (1974) with a few thousand particles and that now reach
billions of particles (70 billions in the Horizon $4\pi$ simulation,
Teyssier et al. 2009), have obtained similar results, with matter also
concentrating along filaments and sheets, clusters of galaxies being
located at the intersection of several filaments.  Although clusters are
believed to form mainly at redshifts $z>1$, they appear still to be
accreting galaxies and groups of galaxies along the filaments they are
connected to, as suggested for example by the analysis of large
spectroscopic surveys from which the dynamical properties of clusters
can be derived (for example for the Coma cluster, based on 715
spectroscopic redshifts, Adami et al. 2009).

Several methods have recently been developed to detect filamentary
patterns in large surveys, such as the Bisous model (Stoica et
al. 2005), an object point process with interactions, whose topology
is based on the Felix model (Shivashankar et al. 2015), or the
FilFinder algorithm that is based on techniques of mathematical
morphology (Koch \& Rosolowsky 2015). The Bisous model has been used
by Tempel et al. (2014a) to build a public catalogue of filaments
based on the SDSS data release~8. The longest filaments in this
catalogue can reach 60~h$^{-1}$~Mpc at a maximum redshift of $z\sim
0.15$. Interesting results were obtained by this team, showing in
particular that galaxies and groups are not uniformly distributed
along filaments, but tend to form a regular pattern with a
characteristic length of between 4~h$^{-1}$~Mpc and 7~h$^{-1}$~Mpc,
depending on the statistics used (Tempel et al. 2014b).  The existence
of a statistically significant alignment between the satellite galaxy
positions and filament axes was also found (Tempel et al. 2015), as
well as an alignment between the orientation of galaxy pairs and their
host filaments (Tempel \& Tamm 2015).

However, the detection of filaments at the intersection of which
clusters are located remains difficult and limited to relatively
low-redshift clusters. At optical or infrared wavelengths, filaments
have been found in a few systems, in particular between two
neighbouring clusters, based on spectroscopic data. In some cases star
formation appears enhanced, as seen between the clusters Abell~1763
and Abell ~1770 ($z=0.23$), where starburst galaxies are numerous
(Edwards et al. 2010). Weak-lensing techniques have also allowed
detecting filaments, such as that between Abell~222 and Abell~223, at
redshift $z=0.21$ (Dietrich et al. 2012).  According to Dietrich and
collaborators, this filament coincides with an overdensity of galaxies
and with diffuse, soft-X-ray emission, and contributes a mass similar
to that of an additional galaxy cluster to the total mass of the
supercluster. In X-rays, the detection of filaments remains difficult,
however, because of their very faint emission (see e.g. the filament
in Abell~85, which was interpreted as due to groups falling onto the
main cluster, Durret et al. 2003, 2005).

In the present paper, we aim to search for filaments of galaxies
around clusters at relatively high redshifts, taking advantage of the
large database that we have collected in the DAFT/FADA survey for
clusters in the $0.4<z<0.9$ redshift range. This survey comprises 90
clusters selected to be massive from their X-ray/ROSAT properties, and
with HST images available in at least two bands. We then obtained deep
imaging data in five optical bands and one infrared band, and as many
galaxy spectroscopic redshifts as possible, either from our own
observations with various telescopes (we were granted about 70 nights
of telescope time altogether) and from archive data. A short
description of the project can be found here:
http://cesam.lam.fr/DAFT/index.php.

We applied a method that consists of selecting galaxies that are
likely to belong to each cluster based on a colour-magnitude diagram,
and of computing density maps for these selected galaxies with an
adaptive kernel method to estimate the extent of the galaxy
distribution around each cluster.  This method is very similar to that
applied by Kartaltepe et al. (2008) on fourteen clusters from the MACS
sample at $z\sim 0.5$. This is an exploratory study to determine
whether we can apply this method to search for extensions and
filaments around a much larger sample of clusters.

The paper is organized as follows. The data and method with which we
searched for filaments and estimated their lengths are described in
Sect.~2, and results are shown in Sect.~3 and discussed in Sect.~4.

We transformed angles into physical units using Ned Wright's Cosmology
Calculator\footnote{http://www.astro.ucla.edu/~wright/CosmoCalc.html}
with H$_0$=70~km~s$^{-1}$~Mpc$^{-1}$, $\Omega _\Lambda = 0.7$, and
$\Omega_M=0.3$.

\section{Data and method}

\subsection{Data}
%voir /home/mencia2NS/durret/FADAS/grande_echelle/Suite_Ana/histomags/completude
\begin{table*}[t!]
  \caption{Clusters in our sample in order of increasing RA. The 
    columns are (1)~cluster name, (2)~RA (J2000.0), (3)~DEC (J2000.0), (4)~cluster redshift,
    scale (in kpc/arcsec), (5)~scale at cluster redshift in kpc/arcsec, (6)~
origin of the data: M for CFHT/Megacam, S 
    for Subaru/SuprimeCam and SC for Subaru/SuprimeCam coming from the CLASH survey,
    (7)~set of filters used, (8)~number of spectroscopic redshifts in the cluster range,
    (9)~magnitude completeness limit in the bluest band, (10)~magnitude completeness limit in the reddest band.
    Two fields contain two clusters each: SEXCLAS 12 and SEXCLAS 13, and CXOSEXSI J205617.1-044155 and MS 2053.7-0449. }
\begin{tabular}{lrrccccrrl}
\hline
\hline
                          &           &          &          &             &            & & & & \\
Cluster name              & RA~~~     & DEC~~    & redshift & scale       & telescope/ & filters & Nz & mag$_{lim,1}$ & mag$_{lim,2}$ \\
                          & (J2000.0) & (J2000.0)&          & (kpc/arcsec)& camera     & & & & \\
                          &           &          &          &             &            & & & & \\
\hline
                          &           &          &        &       &              &     &      & \\
Cl0016+1609               &   4.63888 &  16.4433 & 0.5455 & 6.385 & M & $g', i'$ & 224 & 24.0 & 23.0 \\ 
Cl J0152.7-1357           &  28.17083 & -13.9625 & 0.8310 & 7.603 & S & V, R     & 201 & 25.5 & 25.0 \\ 
RCS J0224-0002            &  36.14320 &  -0.0415 & 0.7730 & 7.419 & M & $g', i'$ &   7 & 24.0 & 22.5 \\ 
PDCS 018                  &  36.85625 &   0.6678 & 0.4000 & 5.373 & M & $g', i'$ &  39 & 25.0 & 23.5 \\ 
XDCS cm J032903.1+025640  &  52.26175 &   2.9403 & 0.4122 & 5.471 & M & $g', i'$ &   1 & 25.0 & 23.5 \\ 
MACS J0454.1-0300         &  73.54552 &  -3.0187 & 0.5377 & 6.339 & M & $g', z'$ & 312 & 24.0 & 21.5 \\ 
MACS J0647.7+7015         & 101.94125 &  70.2508 & 0.5907 & 6.637 &SC & V, I     &   1 & 24.5 & 24.0 \\
MACS J0717+3745           & 109.39083 &  37.7556 & 0.5458 & 6.387 & S & V, I     & 452 & 24.5 & 24.0 \\ 
MACS J0744.9+3927         & 116.21583 &  39.4592 & 0.6860 & 7.087 &SC & V, I     &  72 & 24.5 & 24.0 \\ 
Abell 851                 & 145.73601 &  46.9894 & 0.4069 & 5.429 & M & $g', i'$ & 126 & 24.5 & 23.5 \\ 
SEXCLAS 12                & 163.15917 &  57.5137 & 0.7080 & 7.178 & M & $r', z'$ &  20 & 23.0 & 22.5 \\ 
SEXCLAS 13                & 163.22583 &  57.5360 & 0.6640 & 6.992 & M & $r', z'$ &  20 & 23.0 & 22.5 \\ 
RXC J1206.2-0848          & 181.54991 &  -8.8000 & 0.4400 & 5.685 & M & $r', z'$ &  52 & 24.0 & 22.0 \\ 
BMW-HRI J122657.3+333253  & 186.74167 &  33.5484 & 0.8900 & 7.765 & S & V, I     &   1 & 24.0 & 23.0 \\ 
ZwCl 1332.8+5043          & 203.58333 &  50.5151 & 0.6200 & 6.786 & M & $g', r'$ &   4 & 24.0 & 24.0 \\ 
MJM98 034                 & 203.80742 &  37.8156 & 0.3830 & 5.231 & M & $g', i'$ &  23 & 24.5 & 23.5 \\ 
LCDCS 0829                & 206.88333 & -11.7617 & 0.4510 & 5.767 & M & $g', i'$ &  47 & 24.0 & 22.0 \\ 
3C 295 cluster            & 212.83396 &  52.2025 & 0.4600 & 5.831 & M & $g', i'$ &  29 & 24.0 & 23.5 \\ 
MACS J1423.8+2404         & 215.95125 &  24.0797 & 0.5450 & 6.382 & S & V, I     &   1 & 24.5 & 23.5 \\ 
RX J1524.6+0957           & 231.16792 &   9.9608 & 0.5160 & 6.206 & M & $g', r'$ &   2 & 24.0 & 24.0 \\ 
RCS J1620.2+2929          & 245.05000 &  29.4833 & 0.8700 & 7.713 & M & $g', i'$ &   1 & 24.0 & 22.0 \\ 
MACS J1621.4+3810         & 245.35000 &  38.1672 & 0.4650 & 5.867 & S & V, I     &   2 & 24.5 & 23.5 \\ 
MS 1621.5+2640            & 245.89863 &  26.5638 & 0.4260 & 5.579 & M & $g', i'$ &  77 & 24.0 & 23.0 \\ 
OC02 J1701+6412           & 255.35659 &  64.2368 & 0.4530 & 5.781 & M & $g', i'$ &   5 & 23.5 & 22.0 \\ 
RX J1716.4+6708           & 259.20667 &  67.1417 & 0.8130 & 7.549 & S & V, R     &   4 & 24.5 & 24.5 \\ 
NEP 0200                  & 269.33083 &  66.5253 & 0.6909 & 7.108 & M & $g', i'$ &   4 & 24.5 & 23.5 \\ 
CXOSEXSI J205617.1-044155 & 314.07150 &  -4.6986 & 0.6002 & 6.686 & S/M & V, $z'$  &   1 & 24.5 & 23.0 \\
MS 2053.7-0449            & 314.09321 &  -4.6287 & 0.5830 & 6.596 & S/M & V, $z'$&   1 & 24.5 & 23.0 \\ 
MACS J2129.4-0741         & 322.35833 &  -7.6911 & 0.5889 & 6.627 &SC & V, I     &  48 & 24.0 & 23.0 \\ 
RX J2328.8+1453           & 352.20792 &  14.8867 & 0.4970 & 6.084 & M & $g', i'$ &   2 & 24.5 & 22.5 \\ 
                          &           &          &        &       &   &          &     & & \\
\hline
\end{tabular}
\label{tab:sample}
\end{table*}

The data used for this work come from the DAFT/FADA survey and were
first presented by Guennou et al. (2010) and then analysed by Guennou et
al. (2014, hereafter G14). A list of all the data available for this
survey can be found at http://cesam.lam.fr/DAFT/project.php.  We
limited our analysis to the thirty clusters of the DAFT/FADA survey
for which CFHT/Megacam or Subaru/SuprimeCam wide field data are
available in at least two photometric bands (obtained with the same
telescope, except for MS 2053.7-0449 and CXOSEXSI J205617.1-044155),
either from our own observations or from the archives. The fields of
these two cameras are $1\times 1$~deg$^2$ and $34\times 27$~arcmin$^2$, 
respectively, and only one field per cluster is available.  Whenever
possible, the two photometric bands were chosen to bracket the
4000~\AA\ break.  For the five following qclusters this was not
possible: Cl~J0152.7-1357, BMW-HRI~J122657.3+333253, ZwCl~1332.8+5043,
RX~J1524.6+0957, and RCS~J1620.2+2929.

For three clusters (indicated with SC in the sixth column of Table~1)
the Subaru/SuprimeCam archive images had too short exposure times to
be usable, therefore we retrieved the CLASH
catalogue\footnote{https://archive.stsci.edu/prepds/clash/} and based
our analysis on the CLASH V and I band data.  For MACS~J0717+3745 we
analysed both archive images and the CLASH data and found similar
results. The full cluster list is given in Table~\ref{tab:sample}.  In
the last two columns of this table, we give an estimate of the
completeness limits of the two images from which the cluster galaxies
were selected. For this, we have drawn magnitude histograms in 0.5
magnitude bins. To be conservative, we consider that the histograms
are complete up to two bins below the bin where the counts
flatten. The Subaru data are of course deeper than the Megacam data,
but although our data are not fully homogeneous, they are of
sufficient quality and homogeneity for a pilot study.

\subsection{Method}

\begin{figure}[h!] %fig1
\centering 
%\mbox{\psfig{figure=figure1c.ps,width=7cm,clip=true}}
\includegraphics[angle=0,width=6cm]{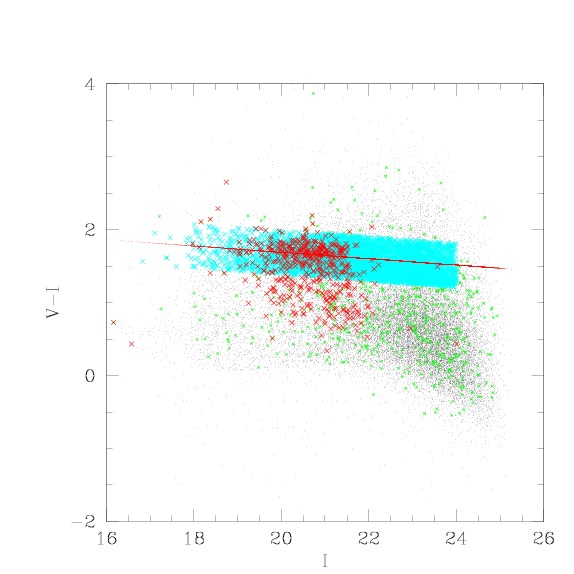}
\caption[]{Colour-magnitude diagram for MACS~J0717+3745. The black
  points are all the galaxies in the field. The green crosses are the
  galaxies in a radius of 1~Mpc around the cluster centre. The red
  crosses show all the galaxies with spectroscopic redshifts in the
  cluster range (taken to be [0.530,0.560]). The red line is the best
  fit of the cluster red sequence.  The cyan points show the galaxies
  with a colour within $\pm 0.3$ of the cluster red sequence, and
  selected as belonging to the cluster. }
\label{fig:coulmag}
\end{figure}

\begin{figure} %fig2
\centering 
\includegraphics[angle=0,width=6cm]{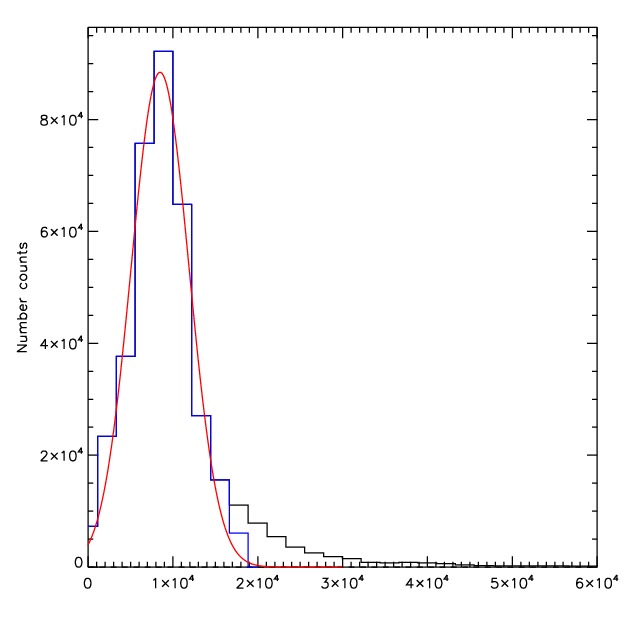}
%\mbox{\psfig{figure=histogramme_MACS0717_clip_FS.eps,width=6cm,angle=0}}
%\mbox{\psfig{figure=histogramme_RX1716_cut_clip.ps,width=6cm,clip=true}}
\caption{Histogram of the pixel intensities for the density map of
  MACS~J0717+3745. The black line shows the full histogram, the blue
  line the histogram obtained after applying a 2.5$\sigma$ clipping
  (see text).  The best fit to the Gaussian noise is shown as a red
  line.}
\label{fig:histobruit}
\end{figure}

To map the galaxy distribution around the clusters of our sample, we
first selected the galaxies that have a high probability to be at the
cluster redshift. For this, we first ran SExtractor (Bertin \& Arnouts
1996) on the images to assemble catalogues with magnitudes in two bands
for each cluster. The magnitudes we used are MAG\_AUTO.  We then
separated galaxies from stars using a maximum surface brightness of the
light profile versus magnitude diagram and eliminated stars. We then
corrected for Galactic extinction using the maps of Schlegel et al. (1998).

We also retrieved all the spectroscopic redshifts available in our own
data and in NED for each cluster. Most of the redshift histograms thus
obtained have been shown by G14, so we only show the redshift
histograms that were not in G14 in Appendix~C.

We then drew colour-magnitude diagrams for each cluster and
superimposed the positions of the galaxies with spectroscopic
redshifts that belong to the cluster.  This allowed defining the red
sequence of the early-type galaxies that belong to the cluster and to
fit it with a linear function. Because the red sequence slope has been
observed to not evolve across the redshift range considered here
(e.g. De Lucia et al. 2007), we assumed a fixed slope of
$-0.0436$. Since for many clusters we only have a few spectroscopic
redshifts, we preferred to take a fixed slope for the red sequence;
this is the same we used in several of our papers (Durret et
al. 2011a, Martinet et al. 2015a), to treat all the clusters in a
consistent way. For the same reason, we also chose to select the same
width of $\pm 0.3$ on either side of the red sequences of all the
clusters. This value is often used in the literature (see
e.g. Lagan\'a et al 2009 and references therein). For the clusters
with many spectroscopic members, this value allows encompassing these
members quite well, as in Fig.~1. These choices are obviously
simplifications, since the red-sequence slope and width are not
exactly the same for all bands, but with the available data we would
add noise by applying a more elaborate selection.  Only galaxies with
magnitudes brighter than $r'\sim 24$ (or the equivalent in other
filters) were taken into account to perform the red sequence fit.  We
then selected all the galaxies within $\pm 0.3$ magnitudes of this
sequence as probable cluster members, as illustrated in
Fig.~\ref{fig:coulmag} for MACS~J0717+3745. For clusters that only
have a few spectroscopic redshifts, we imposed a value of the
intercept by eye, based on the determinations computed for clusters
with many spectroscopic redshifts, and requesting that the intercept
increases with redshift in agreement with the colours given by
Fukugita et al. (1995). We then extracted the catalogue of probable
cluster galaxies in the same way as above.

We thus obtained for each cluster a catalogue of galaxies that are
likely to belong to the cluster (hereafter ``cluster galaxies'').  For
each cluster, we then computed a 2D density map based on the cluster
members selected with the red sequence. To do this, we applied the
adaptive kernel technique with a generalized Epanechnikov kernel as
suggested by Silverman (1986). A summary of our implementation is
given in Dantas et al (1997). It is based on an earlier version
developed by Timothy Beers (ADAPT2) and further improved by Biviano et
al. (1996).  The statistical significance is established by bootstrap
resampling of the data. A density map is computed for each new
realisation of the distribution. For each pixel of the map, the final
value is taken as the mean over all realisations. A mean bootstrapped
map of the distribution is thus obtained (see Mazure et al.,
2007). The number of bootstraps used here is 100.

To derive the significance level of our detections, it is necessary to
estimate the mean value and dispersion of the background of each
image. For this, we drew the histogram of the pixel intensities for
each density map. An example is shown in Fig.~\ref{fig:histobruit} for
MACS~J0717+3745. We applied a 2.5$\sigma$ clipping to eliminate those
pixels of the image that have high values and correspond to objects in
the image. We then redrew the histogram of the pixel intensities after
clipping and fitted this distribution with a Gaussian. For each
cluster, the mean value of the Gaussian will give the mean background
level, and the width of the Gaussian will give the dispersion, which
we call $\sigma$. We then computed the values of the contours that
correspond to 3$\sigma$ detections as the background plus
3$\sigma$. In all the figures of the following Section where density
maps with contours are presented, we show contours starting at
3$\sigma$ and increasing by 1$\sigma$. As mentioned in the previous
subsection, the depths of all the images do not strongly differ from
one cluster to another for a given telescope and filter, therefore the
Megacam density maps on the one hand and the Subaru maps on the other
hand can be considered as roughly homogeneous. Hence, our 3$\sigma$
detections can be considered as roughly homogeneous, at least for a
given telescope and set of filters.

Interestingly, for several clusters for which the red sequence was
defined by two sets of filters (for example $g'$ and $r'$ , and $g'$
and $i'$) the significance level of the detection is higher when the
filter set brackets the 4000~\AA\ break well, but the extent of the
detection on the density map remains roughly the same.  This gives us
confidence on the detections achieved even for filter sets that do not
bracket the 4000~\AA\ break well.

\subsection{Reliability tests of the galaxy selection}

To test the reliability of our density maps, we made several tests on
MACS~J0717+3745, a bright cluster for which hundreds of spectroscopic
redshifts are available. Our best fit is $V-I = -0.0436\times I+b$,
with $b=2.75$. Keeping our selection of galaxies within $\pm 0.3$
magnitudes from the red sequence, we first changed the value of $b$ by
steps of 0.25, and built density maps for the following values: 2.50,
2.75, 3.00, and 3.25. Figure~\ref{fig:0717_varyb} clearly shows for
$b=2.75$ that the cluster and its filament are detected with the
largest extent at $3\sigma$.  The value $b=2.50$ gives similar
results, but with a somewhat smaller extension, while the cluster
fades for $b=3.00$ and 3.25.  Second, we fixed $b=2.75$ and selected
galaxies within $\pm 0.2$ and $\pm 0.4$ from the red sequence.
Results are shown in Fig.~\ref{fig:0717_varyd}. The best compromise is
$\pm 0.3$ because $\pm 0.2$ reduces the cluster extension and $\pm
0.4$ adds noise by making other structures appear on the density map,
which are probably mostly foreground structures.  We therefore
conclude that the parameters chosen to select the ``cluster galaxies''
that were used to compute the density maps are the best choice, and
our detections remain reliable even if these parameters change
slightly. This is important in particular for clusters for which the
red sequence is more difficult to determine because only a few
spectroscopic redshifts are available for them.

\subsection{Reliability tests of the cluster detection levels}

When studying the BMW-HRI J122657.3+333253 cluster, we faced several
problems that led us to test the method we used to estimate the
detection levels of our clusters.

First, our Subaru/SuprimeCam image suffered from contamination by
diffuse light that is due to the bright nearby galaxy NGC~4395, which
is located about 20~arcmin west of the cluster centre. Because this
diffuse light was not masked during the SExtractor processing, a
number of false detections were included in our catalogue, all of them
on the west side of the field, and these spurious objects also
strongly contaminated the cluster red sequence.

We therefore cut the density map to eliminate this structure and
calculated contour levels. After this cut, Fig.~\ref{fig:bmw1226tests}
(top) shows three bright structures close to the image edges, but
BMW-HRI J122657.3+333253 is not detected at the $3\sigma$
level. However, we clearly see the cluster, the position of which is
indicated with a white circle in Fig.~\ref{fig:bmw1226tests}, and it
corresponds to the X-ray detection (see G14, Fig.~A.19).

This might mean that very bright structures artificially increase the
level of the background we compute, and therefore also that of the
$3\sigma$ contours required for an object to be considered as
detected. We therefore made two attempts. First, we cut the density
map to eliminate the three bright sources at the edges of the field
and recomputed the contour levels. With this approach, BMW-HRI
J122657.3+333253 is detected at $4\sigma$, as seen in
Fig.~\ref{fig:bmw1226tests} (middle). Second, we cut the galaxy
catalogue to the same size as the middle figure and recomputed the
density map.  The result is shown in Fig.~\ref{fig:bmw1226tests}
(bottom). Here, the cluster is detected at $15\sigma$ and structures
that were too faint to be clearly visible before are now clearly
detected.  East of the cluster, a structure is detected at the
$5\sigma$ level. It coincides with structure 1 in M15, which can be
identified in NED with the cluster X-class~1808, at redshift
$z=0.766$. These detection levels are to be taken with caution,
however, as the area over which the background has been estimated is
too small to provide reliable statistics.

This shows that very bright sources in the field can artificially
enhance our estimation of the background of the density map and cause
us to underestimate the significance level of the structures that we
detect. In addition, the initial smoothing scale applied to the data
(i.e. the initial kernel dimension) depends on the 2D galaxy
distribution in our sample. Very bright sources in the field tend to
slightly oversmooth our 2D density maps in regions of lower density
(compared to the density of very bright sources, including the cluster
we are trying to detect). As a result, we might lose some details in
the close environment of the cluster if sources brighter than the
cluster are present in the field (as seen in the tests for BMW-HRI
J122657.3+333253).

We therefore made a fourth attempt, which was to trim the catalogue of
cluster galaxies to eliminate NGC~4395 as well as other edge effects
(seen as bright sources in the corners of the density map), and we
recomputed the density map. We consider this last map as the final
result for BMW-HRI J122657.3+333253, as presented in Sect.~3.2.6. and
Fig.~\ref{fig:bmw1226}.  However, we must keep in mind that the
significance levels we report may be underestimated when very bright
structures are present as well. We checked that none of the other
clusters showed such strong edge effects.

As a further test, we considered the cases of the five clusters where
a structure with a brightness similar to or higher than that of the
cluster was present in the density map: ZwCl 1332.8+5043, RCS
J0224-0002, PDCS 018, XDCS~cm J032903.1+025640, and MJM98 034.  We
spatially cut the cluster galaxy catalogues (i.e. the galaxies
selected to belong to each of these clusters) and recomputed the
density maps and contours (as in the bottom panel of
Fig.~\ref{fig:bmw1226tests}). In all cases, the significance levels of
the cluster detections increased by a modest amount ($1\sigma$) and
the cluster extents became only slightly larger. This means that the
contamination only becomes important when an object with a brightness
higher than that of the cluster is present.

\section{Results}

We present our main results in this section. They are based on the
density map of each cluster, onto which we superimposed the intensity
contours computed as explained above, starting at $3\sigma$.  We only
show those parts of the images in the figures that contain
information. However, the background was always computed on the entire
images. To indicate the scale, we plot a circle of 1~Mpc radius in all
figures. This is centred on the position of the cluster taken from NED
and given in Table~1.  In some cases the circle is not exactly centred
on the maximum of the galaxy density map. This is probably because the
cluster position given by NED can have different origins (the
brightest cluster galaxy, the X-ray maximum, or the centroid of the
X-ray emission, etc.). We briefly discuss this question for each
object when this is the case.

Figures~\ref{fig:cl0016} to \ref{fig:ms2053} show the galaxy density
maps. Yellow denotes the highest density, purple the lowest, and dark
blue corresponds to no galaxies. These figures are illustrative, and
the scales and density levels are not the same in each figure.

For each cluster, an approximate size of the extension was obtained by
adjusting by eye an ellipse that fits the $3\sigma$ contours best.
The major and minor axes of the ellipses are given in
Table~\ref{tab:ellipses} for the extended objects. Two ellipses were
necessary for MACS J0717+3745, but for all the other objects, a single
ellipse was sufficient.

Whenever possible, we compare our results with the mass distributions
derived by Martinet et al. (2015b, hereafter M15) with a weak-lensing
analysis, and in some cases also from X-rays. These three methods are
complementary. For a given cluster, the density map at the cluster
redshift is sensitive to clusters and filaments in the plane of the
sky.  On the other hand, weak lensing detects mass along the line of
sight, and therefore clusters and filaments regardless of their
orientation, but it can be contaminated by projection effects of
foreground and background structures. Finally, X-rays trace the hot
baryons of groups and clusters.

We separate our results into four different subsections: the twelve
clusters showing an extended structure and/or filaments detected at a
$3\sigma$ level, the eleven clusters with neighbouring structures but
showing no large elongation or filament detected at a $3\sigma$ level,
the three clusters showing no large extension and no significantly
detected neighbouring structures or filaments, and the four clusters
whose contrast is too low to be detected at $3\sigma$ on the density
maps.

\subsection{Twelve clusters showing an extended structure 
and/or filaments}

\begin{table}[t!]
  \caption{Sizes of the major (a) and minor (b) axes of the ellipses 
    (in Mpc) that fit the elongations for all the clusters in our sample 
    that show an 
    extended and/or elongated structure (the yellow ellipses in the figures;
    for MACS J0717+3745 there are two 
    ellipses, as illustrated in Fig.~\ref{fig:macs0717}). }
\begin{tabular}{lrr}
\hline
\hline
                          &          & \\
Cluster name              & a (Mpc)  & b (Mpc) \\
                          &          & \\
\hline
                          &          & \\
Cl0016+1609               & 7.4 & 3.2 \\ 
                          & 4.8 & 3.4    \\
MACS J0647.7+7015         & 6.8 & 2.2  \\
MACS J0717+3745           & 6.0 & 1.8 \\ 
                          & 3.2 & 2.1 \\
MACS J0744.9+3927         & 3.8 & 1.5  \\
RXC J1206.2-0848          & 5.7 & 2.4 \\  
ZwCl 1332.8+5043          & 5.8 & 5.4 \\  
LCDCS 0829                & 7.5 & 3.3 \\  
MACS J1423.8+2404         & 6.0 & 3.0 \\  
MACS J1621.4+3810         & 7.6 & 2.1 \\ 
MS 1621.5+2640            & 6.0 & 3.8 \\  
RX J1716.4+6708           & 3.5 & 1.1 \\  
MACS~J2129.4-0741         & 3.7 & 1.6 \\
                          &          & \\
\hline
\end{tabular}
\label{tab:ellipses}
\end{table}

We first present here the twelve clusters in which we detect an
extended or elongated structure and/or filaments.

\subsubsection{Cl~0016+1609 (z=0.5455)} %YES 4.6 fig.3

\begin{figure} 
\centering 
%\mbox{\psfig{figure=Cl0016gi_cont_ell_zspec.ps,width=7cm,clip=true}}
%\includegraphics[angle=0,width=6cm]{Cl0016gi_cont_ell_zspec.jpg}
\includegraphics[angle=0,width=6cm]{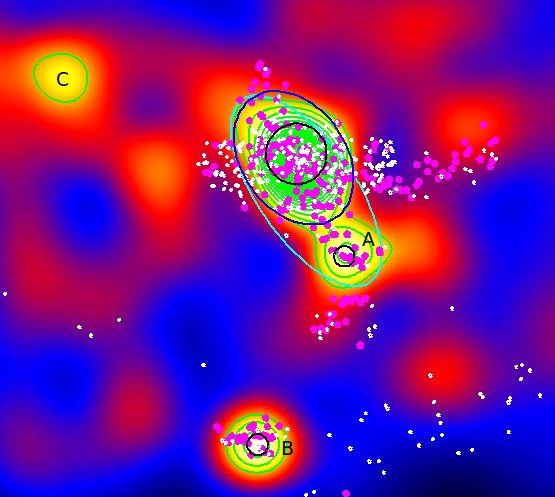}
\caption{Density map of the galaxies selected as belonging to the
  cluster Cl~0016+1609 (z=0.5455).  The large black circle (in some of
  the following figures it may be drawn in other colours to be more
  clearly visible) is centred on the position of the cluster given in
  Table~1 and has a radius of 1~Mpc, as in all following figures.  The
  green contour levels start at $3\sigma$ and increase by $1\sigma$.
  The cyan ellipse (often drawn in yellow in other figures) indicates
  the maximum extent of the $3\sigma$ contours, with the values of the
  ellipse major and minor axes given in Table~\ref{tab:ellipses}. The
  dark blue ellipse indicates the maximum extent if the neighbouring
  cluster A is excluded. The small white points show the positions
  of the galaxies with a measured spectroscopic redshift, and the
  magenta points those that are in the approximate cluster redshift
  range (given in the figure captions, here $0.53<z<0.57$).  The
  medium-size circles (here black, but also blue or white in some of
  the following figures) indicate clusters that we detect and that can
  be identified with a galaxy cluster or group in NED. North is up and
  east is left in all figures.  }
\label{fig:cl0016}
\end{figure}

Cl0016+1609 has an elongated structure at least 4.8~Mpc in length
(Fig.~\ref{fig:cl0016}) and several possible companion clusters.  The
cluster RXJ0018.3+1618 (at $z=0.5506$) found in NED and shown as a
small black circle labelled A is detected at $5\sigma$. It is probably
a low-mass companion of Cl0016. It is located 9.8~arcmin (3.8~Mpc at
the cluster redshift) south-west of Cl0016 and is also mentioned by
\kart . The total extent of the system including Cl0016+1609 and
cluster A is 7.4~Mpc (see Table~\ref{tab:ellipses}). A source labelled
B is detected at $5\sigma$ 25.3~arcmin (9.7~Mpc) south of Cl0016 and
can be identified in NED with the cluster RX~J0018.8+1602 at
$z=0.5406$.  Higuchi et al. (2015) calculated the virial masses for
Cl0016+1609 and RX~J0018.8+1602 and found respective values of
$(20.9\pm 2.1)\times 10^{14}$~M$_\odot$ and $(4.5\pm 1.1)\times
10^{14}$~M$_\odot$.  The source labelled C detected at $3\sigma$
north-north-east of Cl0016 has no identification in NED. There are 292
spectroscopic redshifts in the $0.53<z<0.57$ range; their distribution
is roughly Gaussian.

\subsubsection{MACS J0647.7+7015  (z=0.5907)} %YES 101.9 fig4

\begin{figure}[h!]
\centering 
\includegraphics[angle=0,width=6cm]{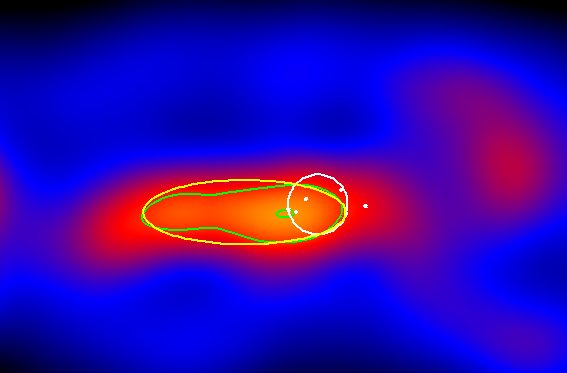}
%\mbox{\psfig{figure=MACS0647_CLASH_cont_ell_zspec.ps,width=7cm,clip=true}}
\caption{Same as Fig.~\ref{fig:cl0016} for MACS J0647.7+7015  (z=0.5907).
}
\label{fig:macs0647}
\end{figure}

MACS J0647.7+7015 appears elongated (Fig.~\ref{fig:macs0647}) but only
at the $3\sigma$ level. We can note, however, that \kart\ detected a
rather large elongation for this object.

\subsubsection{MACS J0717+3745 (z=0.5458) } %YES 109.4 fig5

\begin{figure}[h!]
\centering 
\includegraphics[angle=0,width=6cm]{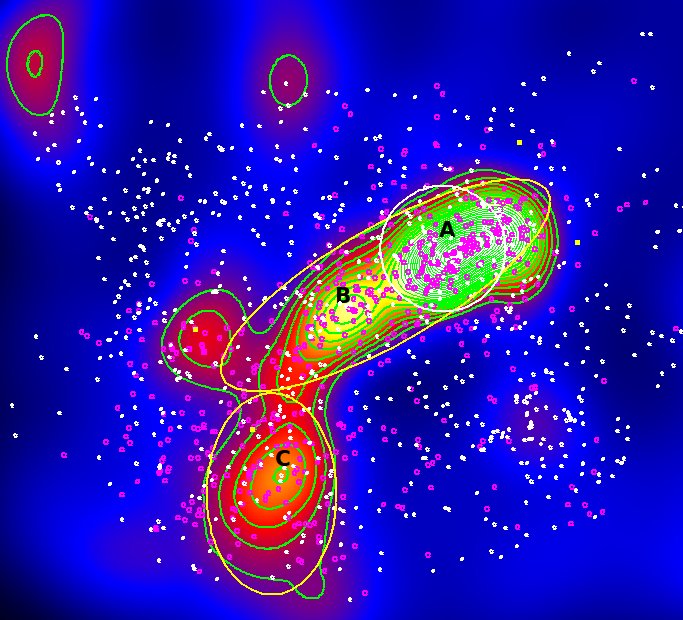}
%\mbox{\psfig{figure=MACS0717_cont_ell_zspec2.ps,width=7cm,clip=true}}
\caption{Same as Fig.~\ref{fig:cl0016} for MACS J0717+3745 (z=0.5458). 
The magenta points correspond to the galaxies with spectroscopic redshifts
in the $0.53<z<0.565$ interval.}
\label{fig:macs0717}
\end{figure}

MACS J0717+3745 is a very massive cluster ($M_{200}^{NFW}=2.26\times
10^{15}$~M$_\odot$ according to M15) that is known to be an extreme
merger in X-rays (Mann \& Ebeling 2012).  Figure~\ref{fig:macs0717}
shows a first elongation of about $6.0\times 1.8$~Mpc that contains
the main cluster labelled A (structure~1 in M15) and a second massive
structure labelled B (structure~2 in M15) to the south-east. A second
elongation of about $3.2\times 2.1$~Mpc (labelled C and coinciding
with structure~3 in M15) is detected farther south.
Figure~\ref{fig:macs0717} is similar to Fig.~2 of \kart.  As discussed
by M15, neither of these elongations is strongly detected in X-rays,
which means that they are probably formed by one or two filaments and
not only by clusters merging at large scales. We note that this
filament was previously reported by Jauzac et al. (2012) and
Medezinski et al. (2013) and is the longest and strongest filament
detected in our sample of clusters.  There are 577 spectroscopic
redshifts in the $0.530<z<0.565$ range, the histogram of their
distribution is shown in Fig.~\ref{fig:histozMACS0717}.

\subsubsection{MACS~J0744.9+3927 (z=0.6860)  } %YES 116.2 fig6

\begin{figure}[h!]
\centering 
\includegraphics[angle=0,width=6cm]{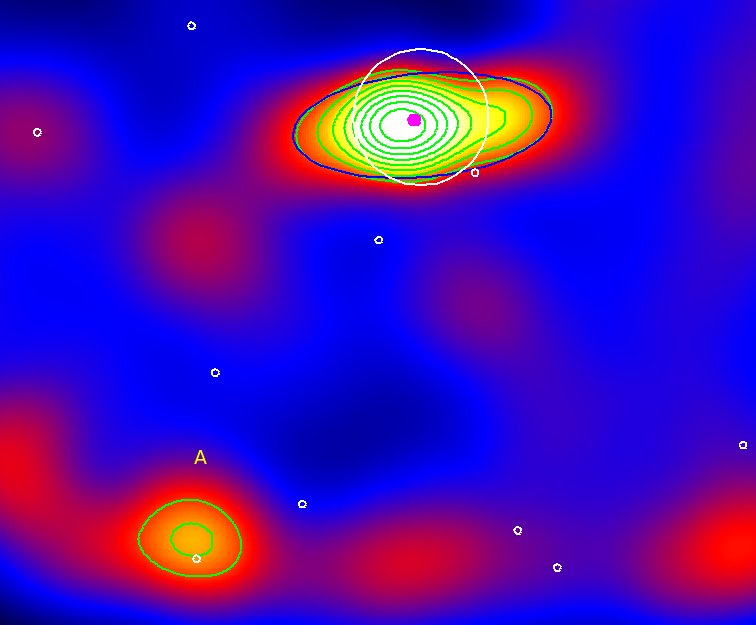}
%\mbox{\psfig{figure=MACS0744_CLASH_cont_ell_zspec.ps,width=7cm,clip=true}}
\caption{Same as Fig.~\ref{fig:cl0016} for MACS J0744.9+3927 (z=0.6860).
The magenta points correspond to the galaxies with spectroscopic redshifts
in the $0.695<z<0.705$ interval.}
\label{fig:macs0744}
\end{figure}

MACS~J0744.9+3927 (z=0.6860) is detected at the $10\sigma$ level and
shows an elongation reaching almost 4~Mpc in the east-west direction
(Fig.~\ref{fig:macs0744}).  This cluster has a high mass, as derived
both from X-rays ($M_{r500}=9.9\times 10^{14}$~M$_\odot$ (G14)) and
from weak lensing ($M_{r<1.5\ Mpc}=20.5\times 10^{14}$~M$_\odot$
(Applegate et al. 2014)). Only two spectroscopic redshifts are
available in the $0.695<z<0.705$ range.

\subsubsection{RXC J1206.2-0848  (z=0.4400)} %YES 181.5 fig7

\begin{figure}[h!]
\centering 
\includegraphics[angle=0,width=6cm]{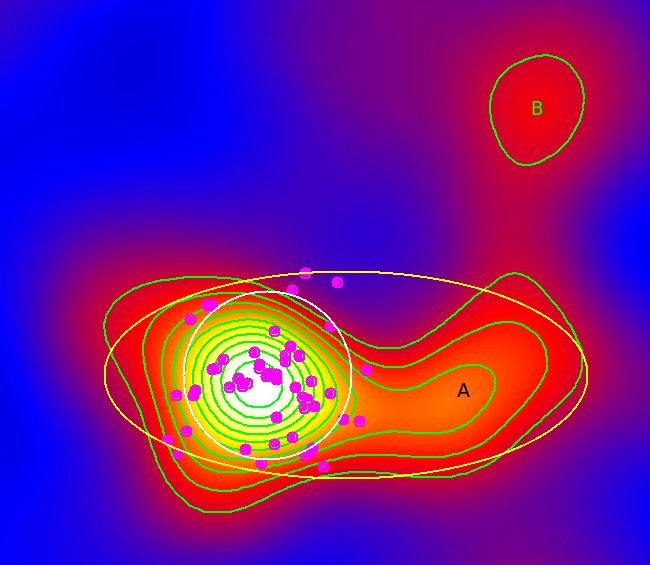}
%\mbox{\psfig{figure=RX1206_cont_ell_zspec.ps,width=7cm,clip=true}}
\caption{Same as Fig.~\ref{fig:cl0016} for RXC J1206.2-0848 (z=0.4400).
The magenta points correspond to the galaxies with spectroscopic redshifts
in the $0.42<z<0.46$ interval.}
\label{fig:rxc1206}
\end{figure}

RXC J1206.2-0848 shows an elongation towards the west (labelled A)
that may be a second less massive cluster (see
Fig.~\ref{fig:rxc1206}). A fainter structure (labelled B) is detected
at $3\sigma$ north of the second one and seems linked to it by a
filament of galaxies, but the detection level of this possible
filament is below $3\sigma$. We were unable to identify these two
structures with clusters in NED.  Several other features are detected
in the field, but barely at a $3\sigma$ level, therefore we only plot
these structures in Fig.~\ref{fig:rxc1206}.  Fifty-two spectroscopic
redshifts are available in the $0.42<z<0.46$ interval, but none in the
filament.

\subsubsection{ZwCl 1332.8+5043 (z=0.6200)} %YES 203.58 fig8

\begin{figure} 
\centering 
\includegraphics[angle=0,width=6cm]{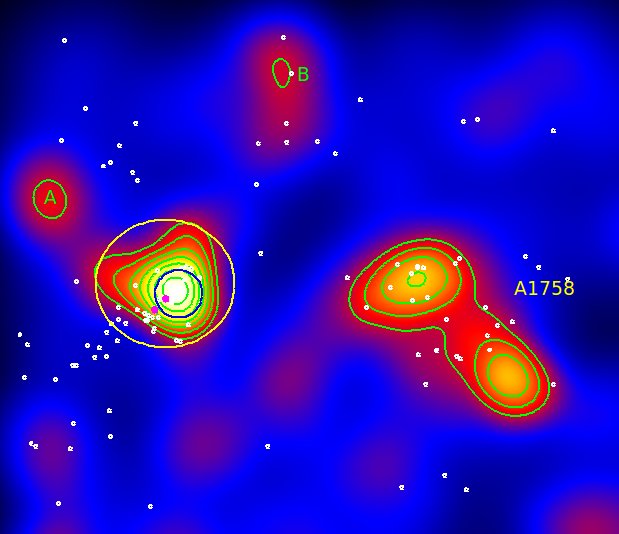}
%\mbox{\psfig{figure=Zw1332bcut_cont_ell_zspec.ps,width=7cm,clip=true}}
\caption{Same as Fig.~\ref{fig:cl0016} for ZwCl 1332.8+5043 (z=0.6200).}
\label{fig:zw1332}
\end{figure}

ZwCl 1332.8+5043 is detected at a $9\sigma$ level in the density map
(Fig.~\ref{fig:zw1332}). Matter may be present between the cluster and
the structure labelled A detected at $3\sigma$ to the north-east, but
its detection level is below $3\sigma$ and there is no identification
in NED for this north-eastern structure. Neither is there a NED
identification for the structure labelled B.  The two features seen
west of the cluster can be identified with the double cluster
Abell~1758 North and South at $z=0.279$. We give in
Table~\ref{tab:GTCZw} the new redshifts measured with the GTC and
OSIRIS in this field and the redshift histogram obtained with these
new redshifts and those found in NED in
Fig.~\ref{fig:histozZw1332}. Unfortunately, only a few discordant
spectroscopic redshifts are available in this zone.

\subsubsection{LCDCS 0829 (z=0.4510)} %YES 206.8 fig9

\begin{figure}[h!]
\centering 
\includegraphics[angle=0,width=6cm]{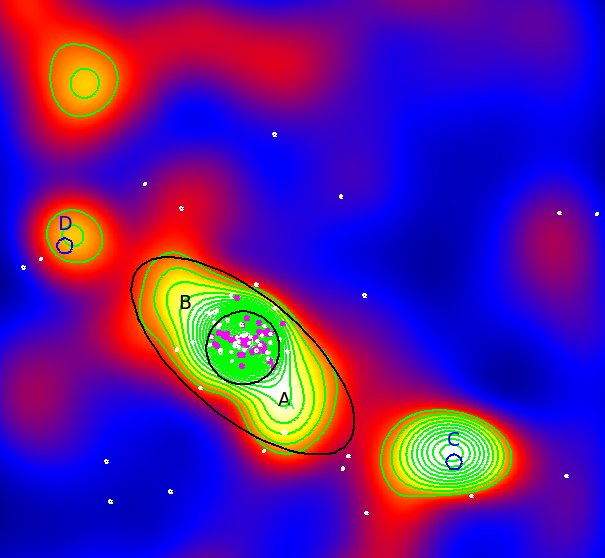}
%\mbox{\psfig{figure=LCDCS0829gi_cont_ell_zspec.ps,width=7cm,clip=true}}
\caption{Same as Fig.~\ref{fig:cl0016} for LCDCS 0829 (z=0.4510).
The magenta points correspond to the galaxies with spectroscopic redshifts
in the $0.44<z<0.46$ interval.}
\label{fig:lcdcs0829}
\end{figure}

LCDCS~0829 (often known as RX~J1347.5-1145) is a very hot and quite
massive cluster ($M_{200}^{NFW}=9.1\times 10^{14}$~M$_\odot$ in M15)
that shows two bright substructures north and south of the cluster in
X-rays (G14).  These structures more or less coincide with those
labelled A and B within the black ellipse of Fig.~\ref{fig:lcdcs0829}.
A third structure (labelled C and coinciding with structure 2 in M15),
much brighter than B, is detected 18~arcmin to the south-west and can
be identified with LCDCS~0825 at redshift $z=0.3900$. A fourth
structure (labelled D) north-east of LCDCS0829 is detected at $4\sigma$
and can be identified with LCDCS~0833, at $z=0.5500$.  The differences
in redshift suggest no link between LCDCS~0829 and its neighbours in
projection on the sky. The redshift histogram of the 43 galaxies with
$0.44<z<0.46$ in LCDCS~0829 is roughly Gaussian and suggests that this
cluster cannot be far from dynamical equilibrium.

\subsubsection{MACS J1423.8+2404 (z=0.5450)} %YES 215.9 fig10

\begin{figure}[h!]
\centering 
\includegraphics[angle=0,width=6cm]{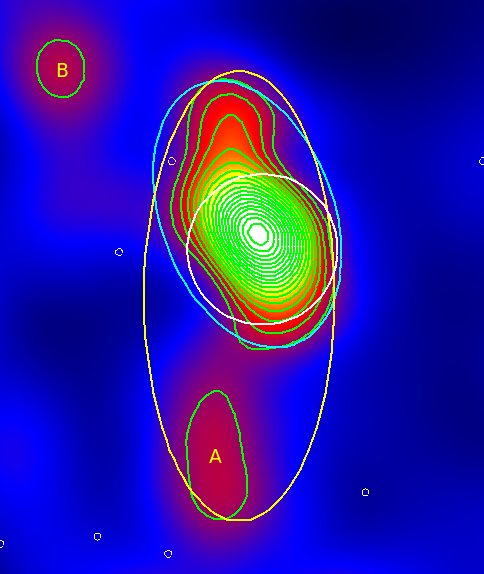}
%\mbox{\psfig{figure=MACS1423_cont_ell_zspec.ps,width=7cm,clip=true}}
\caption{Same as Fig.~\ref{fig:cl0016} for MACS J1423.8+2404
  (z=0.5450).  The yellow ellipse shows the maximum extension
  including the southern structure (about 6~Mpc), the cyan ellipse
  the extension without the southern structure (about 3.7~Mpc).}
\label{fig:macs1423}
\end{figure}

MACS~J1423.8+2404 is a rather massive cluster
($M_{200}^{NFW}=8.2\times 10^{14}$~M$_\odot$ in M15) showing
elongations to the north and south (the cyan ellipse on
Fig.~\ref{fig:macs1423}) that coincide with those detected in X-rays by
G14. It also shows a large structure to the south labelled A within the
yellow ellipse in Fig.~\ref{fig:macs1423} and a fainter one about
9~arcmin to the north-east, labelled B. Both are only detected at
$3\sigma$ and are not detected by \karp ). If the southern structure is
indeed linked to the main cluster, the total elongation is about
$6.0\times 3.0$~Mpc$^2$ (the yellow ellipse), if this is not
the case, it is only
$3.7\times 2.3$~Mpc$^2$ (the cyan ellipse). No redshift close to that
of the cluster is available.

\subsubsection{MACS J1621.4+3810 (z=0.4650)} %YES 245.35 fig11

\begin{figure}[h!]
\centering 
\includegraphics[angle=0,width=6cm]{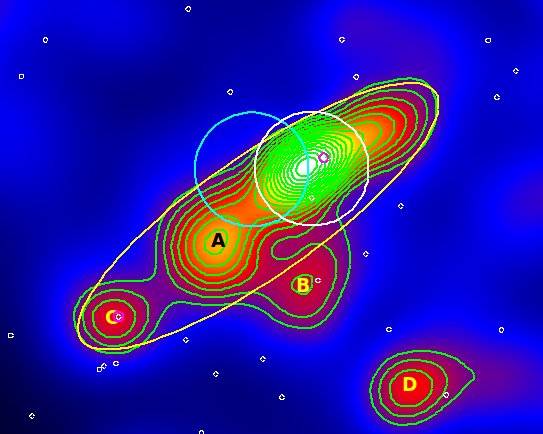}
%\mbox{\psfig{figure=MACS1621_cont_ell_zspec.ps,width=7cm,clip=true}}
\caption{Same as Fig.~\ref{fig:cl0016} for MACS J1621.4+3810
  (z=0.4650).  The cyan circle shows the position of
  MACS~J1621.6+3810. The small magenta circle corresponds to the only
  redshift in the cluster range ($0.44<z<0.49$).}
\label{fig:macs1621}
\end{figure}

MACS~J1621.4+3810 is a moderately massive cluster
($M_{200}^{NFW}=6.4\times 10^{14}$~M$_\odot$ in M15) that appears to
be embedded in a multiple structure at least 7.6~Mpc long along a
north-west -- south-east direction (the yellow ellipse in
Fig.~\ref{fig:macs1621}), including structures A and C (C coincides
with structure 3 in M15). MACS~J1621.6+3810 is also part of this
structure, but does not appear clearly, probably because it is not
very massive. Two other structures are visible south (B) and
south-west (D) of the main cluster.  Although no identifications with
clusters can be made from NED, this system could be a supercluster and
deserves a detailed dynamical analysis. This would require new data,
since no redshift close to that of the cluster is available at
present.  Only three spectroscopic redshifts are available.

\subsubsection{MS 1621.5+2640 (z=0.4260)} %YES 245.9 fig12

\begin{figure}[h!]
\centering 
\includegraphics[angle=0,width=6cm]{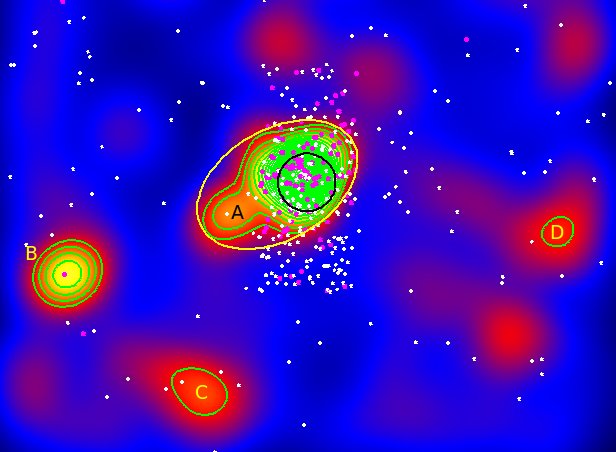}
%\mbox{\psfig{figure=MS1621gi_cont_ell_zspec.ps,width=7cm,clip=true}}
\caption{Same as Fig.~\ref{fig:cl0016} for MS 1621.5+2640 (z=0.4260).
The magenta points correspond to the galaxies with spectroscopic redshifts
in the $0.415<z<0.44$ interval.}
\label{fig:ms1621}
\end{figure}

MS~1621.5+2640 is a rather massive cluster ($M_{200}^{NFW}=9.4\times
10^{14}$~M$_\odot$ in M15) with a much less massive companion
structure located to the south-east (labelled A, within the yellow
ellipse in Fig.~\ref{fig:ms1621}), which is not detected in the
weak-lensing map of M15. Several other structures are visible on the
density map: a $6\sigma$ detection to the south-east (labelled B) that
can be identified with FSVS\_CL J162412+263008 at $z=0.370 $, and two
$3\sigma$ detections to the south and west (labelled C and D) with no
NED identifications. There are 132 spectroscopic redshifts in the
$0.415<z<0.44$ interval, and the redshift histogram appears to be
roughly Gaussian, but there is no spectroscopic information on object
A (within the yellow ellipse of Fig.~\ref{fig:ms1621}).

\subsubsection{RX J1716.4+6708 (z=0.8130)} %259.2 fig13

\begin{figure}[h!]
\centering 
\includegraphics[angle=0,width=6cm]{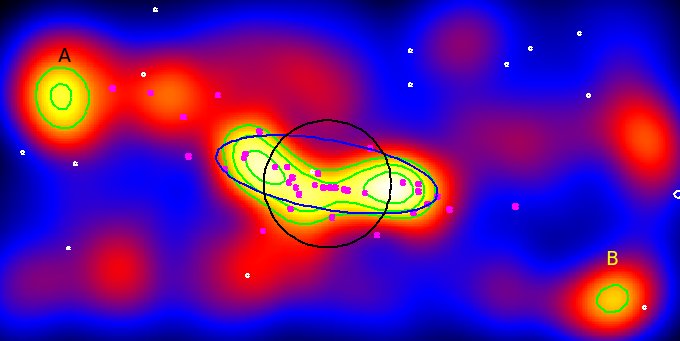}
%\mbox{\psfig{figure=RX1716_cont_ell_zspec.ps,width=7cm,clip=true}}
\caption{Same as Fig.~\ref{fig:cl0016} for RX J1716.4+6708 (z=0.8130).
The magenta points correspond to the galaxies with spectroscopic redshifts
in the $0.79<z<0.83$ interval.}
\label{fig:rx1716}
\end{figure}

RX~J1716.4+6708 is a very massive cluster ($M_{200}^{NFW}=1.46\times
10^{15}$~M$_\odot$ in M15) showing a double structure elongated
roughly in the east--west direction (Fig.~\ref{fig:rx1716}). A third
structure labelled A is detected 9.7~arcmin (4.4~Mpc at the cluster
redshift) towards east-north-east.  Galaxies may be present between
these structures, but their detection does not reach $3\sigma$. A
fourth structure, labelled B, is located 10.7~arcmin (4.9~Mpc at the
cluster redshift) to the south-west.  Thirty-seven spectroscopic
redshifts are available in the cluster range ($0.79<z<0.83$), and their
histogram is shown in Fig.~\ref{fig:histozRX1716}.  The flat
spectroscopic redshift histogram implies that there are at least two
dynamical systems (or more), as confirmed by a Serna-Gerbal analysis
(Serna \& Gerbal 1996), which implies the existence of merging events
occurring in a plane different from that of the sky (G14). It would be
interesting to analyse this system in detail from a dynamical point of
view with more spectroscopic redshifts, since we might be witnessing a
protocluster still in the process of forming, as suggested by Henry et
al. (1997).

\subsubsection{ MACS J2129.4-0741 (z=0.5889)} %259.2 fig14

\begin{figure}[h!]
\centering 
\includegraphics[angle=0,width=6cm]{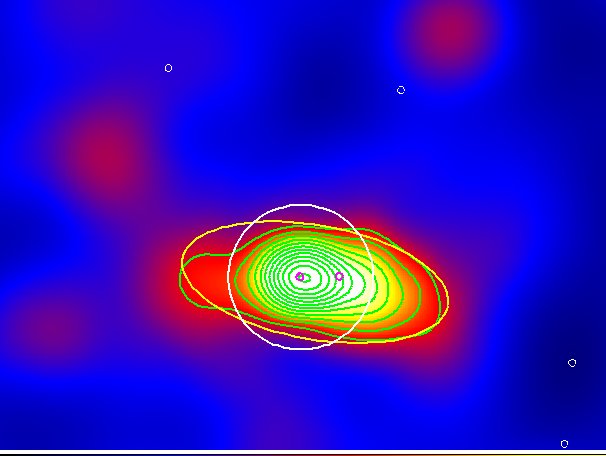}
%\mbox{\psfig{figure=MACS2129_CLASH_cont_ell_zspec.ps,width=7cm,clip=true}}
\caption{Same as Fig.~\ref{fig:cl0016} for MACS~J2129.4-0741 (z=0.5889).
The magenta points correspond to the galaxies with spectroscopic redshifts
in the $0.58<z<0.59$ interval.}
\label{fig:macs2129}
\end{figure}

For MACS J2129.4-0741 the density map shows a strong $14\sigma$
detection, and its structure seen in Fig.~\ref{fig:macs2129} is quite
similar to that found by \kart. The cluster is clearly asymmetric and
elongated roughly along the east-west direction, possibly due to a
merger.

This cluster must be of rather low mass, since Applegate et al. (2014)
were unable to determine its mass through weak lensing. There is no other
source detected at $3\sigma$ in the density map, and only two
spectroscopic redshifts are available in the cluster range.

\subsection{Eleven clusters with neighbouring structures, but without
 extensions or filaments detected at a $3\sigma$ level}

We now present the eleven clusters that have neighbouring structures,
but no filaments or large extensions detected at a $3\sigma$ level.

\subsubsection{Cl J0152.7-1357 (z=0.831)} %YES 28.2 fig15

\begin{figure}[h!]
\centering 
\includegraphics[angle=0,width=6cm]{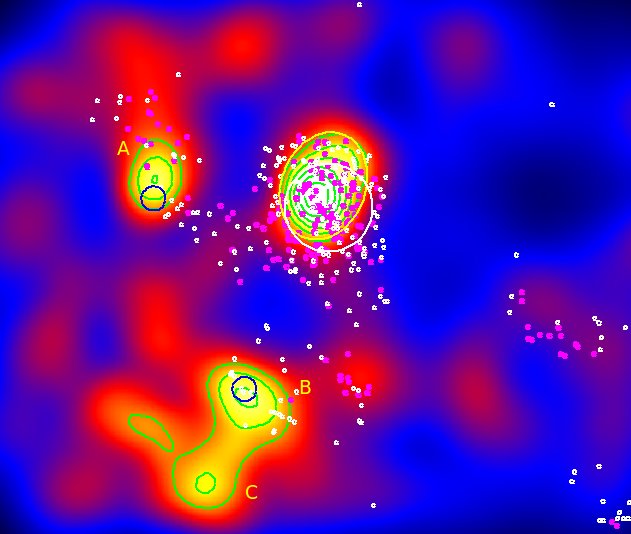}
%\mbox{\psfig{figure=Cl0152_cont_ell_zspec.ps,width=7cm,clip=true}}
\caption{Same as Fig.~\ref{fig:cl0016} for Cl J0152.7-1357 (z=0.831).
The magenta points correspond to the galaxies with spectroscopic redshifts
in the $0.815<z<0.86$ interval.
%The two black circles indicate the structures identified in NED.  
}
\label{fig:cl0152}
\end{figure}

Cl J0152.7-1357 is a very massive cluster ($M_{200}^{NFW}=1.43\times
10^{15}$~M$_\odot$ in M15) known to include two structures that
are in the
process of merging (G14, and references therein), within the 1~Mpc
radius circle of Fig.~\ref{fig:cl0152}. North-north-west of this
circle lies an elongation (the yellow ellipse) that suggests
another possible merger at this scale.  Several other
structures are visible in Fig.~\ref{fig:cl0152}.  The structure
(labelled A) 8.8~arcmin to the east can be identified in NED with the
cluster [BGV2006]~008 at $z=0.577$, while the structure (labelled B,
and corresponding to structure 4 in M15) 10.0~arcmin to the south-east
coincides in position with X-CLASS~0442 at $z=0.7450$.  Since both are
foreground clusters, there is no supercluster in the region of
Cl~J0152.7-1357. The structure labelled C has no counterpart in NED.
There are 201 galaxies with spectroscopic redshifts in the
$0.815<z<0.86$ range.  The corresponding redshift histogram is roughly
bimodal, with 81 galaxies in the $0.815<z\leq 0.835$ interval and 120
galaxies in the $0.835<z<0.86$ range. The positions of these two
systems appear superimposed on the sky, implying that the merger is
taking place more or less in the plane of the sky (also see G14 and
Girardi et al. 2005).

\subsubsection{RCS J0224-0002 (z=0.7730)} %YES 36.1 fig16

\begin{figure} 
\centering 
\includegraphics[angle=0,width=6cm]{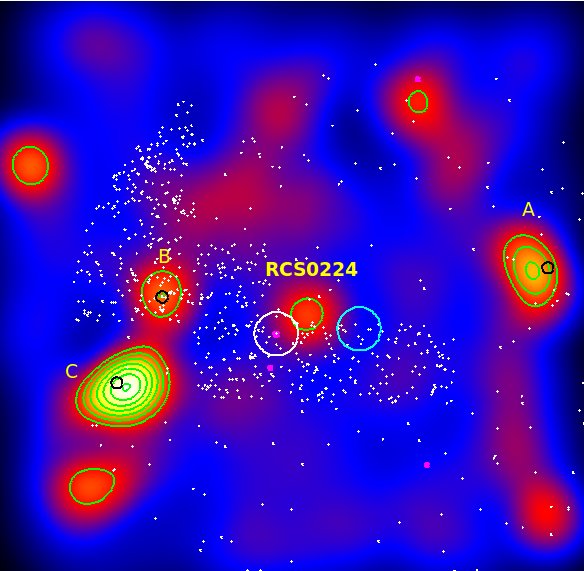}
%\mbox{\psfig{figure=RCS0224_cont_ell_zspec.ps,width=7cm,clip=true}}
\caption{Same as Fig.~\ref{fig:cl0016} for RCS J0224-0002 (z=0.773).
  The cyan and white circles have a radius of 1~Mpc and are
  located at the positions given by Gladders et al. (2002) and by our
  Table~1 (see text), respectively.
The magenta points correspond to the galaxies with spectroscopic redshifts
in the $0.765<z<0.79$ interval.}
\label{fig:rcs0224}
\end{figure}

RCS J0224-0002 (z=0.7730) was discovered by Gladders et al. (2002) to
be a massive cluster showing a system of gravitational arcs.  We note
that the position given by NED corresponds to the name given by
Gladders et al. (2002) with truncated coordinates, 02 24, -00 02,
while on the Megacam image the centre of the cluster is roughly
located at RA=02$^h$24$^{mn}$34.368$^s$, DEC=$-00^\circ$02'29.4''. We
indicate the latter position in Table~\ref{tab:sample} and show both
positions in Fig.~\ref{fig:rcs0224}.  We only detect RCS J0224-0002 at
the $3\sigma$ level (between the two cluster positions mentioned
above), probably because it is quite distant and the exposure times of
our images are rather short (1492~s in $g'$ and 660~s in $i'$).  The
structures labelled A, B, and C can be identified with clusters in NED
that have photometric redshifts in the range 0.41-0.45 and therefore
cannot be in the vicinity of RCS J0224-0002.  There are only five
spectroscopic redshifts in the $0.77<z<0.78$ range.

\subsubsection{PDCS 018 (z=0.4000)} %YES 36.8 fig17

\begin{figure} 
\centering 
\includegraphics[angle=0,width=6cm]{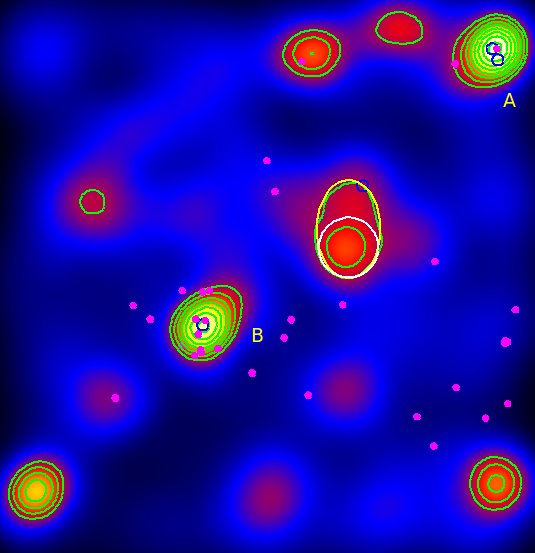}
%\mbox{\psfig{figure=PDCS18_cont_ell_zspec.ps,width=7cm,clip=true}}
\caption{Same as Fig.~\ref{fig:cl0016} for PDCS 018 (z=0.4000).
The magenta points correspond to the galaxies with spectroscopic redshifts
in the $0.392<z<0.405$ interval. }
\label{fig:pdcs018}
\end{figure}

PDCS 018 is a cluster for which very little information is available:
it has only three references in NED, all from the Palomar Distant
Cluster Survey (Postman et al. 1996). In Fig.~\ref{fig:pdcs018},
PDCS~018 appears to be elongated in the north-south direction, possibly due
to an ongoing merger, but in the absence of X-ray data it is difficult
to make a definite statement.  Two bright structures are seen in the
density map 25.2~arcmin north-west (labelled A) and 17.1~arcmin
south-east (labelled B) of the cluster.  The north-western structure
coincides with two clusters in NED: NSCS J022635+010002 at
$z=0.367481$ and VGCF~30 at $z=0.39790$.  The latter structure would
be located 8.1~Mpc from PDCS~018 since they are at the same
redshift. The south-eastern structure can be identified with WHL
J022825.9+003202 at photo$-z=0.414$. The other detected structures
have no identification with groups or clusters in NED. Thirty-four
spectroscopic redshifts are available in the [0.393,0.404] range in a
30~arcmin radius zone. None coincides with the cluster position (given
by NED), but a number of redshifts in the cluster redshift range are
visible throughout Fig.~\ref{fig:pdcs018} (the pink points),
suggesting that we are indeed seeing the large-scale structure in this
region.  The spectroscopic redshift histogram is shown in
Fig.~\ref{fig:histozPDCS18}.  We may therefore be observing a system
of several clusters of which PDCS~018 is not the brightest.

\subsubsection{XDCS cm J032903.1+025640 (z=0.4122)} %YES 52.2 fig18

\begin{figure}[h!]
\centering 
\includegraphics[angle=0,width=6cm]{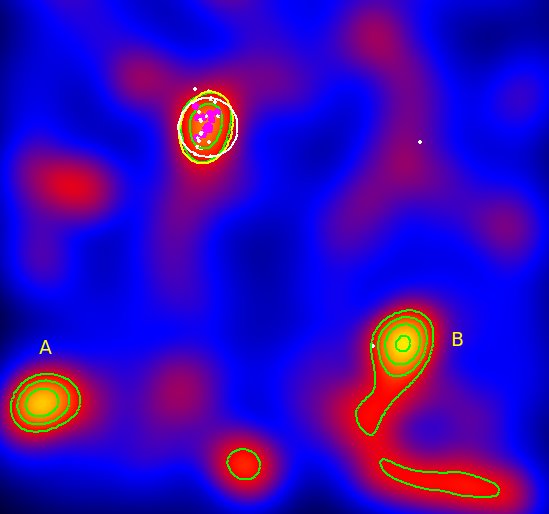}
%\mbox{\psfig{figure=XDCS0329_cont_ell_zspec.ps,width=7cm,clip=true}}
\caption{Same as Fig.~\ref{fig:cl0016} for XDCS cm J032903.1+025640
  (z=0.4122).
The magenta points correspond to the galaxies with spectroscopic redshifts
in the $0.408<z<0.416$ interval.}
\label{fig:xdcs0329}
\end{figure}

XDCS~cm~J032903.1+025640 is a low-mass cluster according to G14
(M$\sim 3\times 10^{14}$~M$_\odot$). Figure~\ref{fig:xdcs0329} suggests
that it could be embedded in a network of structures and/or filaments,
with at least four other structures detected at a $3\sigma$ level and
above. However, there is no identification in NED for the two bright
structures south-east (A) and south-west (B) of the cluster, and the filaments
are too faint to be characterised accurately. Only 13 spectroscopic
redshifts are available in the $0.408<z<0.416$ range, all within the
1~Mpc radius circle.

\subsubsection{ Abell 851 (z=0.4069)} %YES 145.7 fig19

\begin{figure}[h!]
\centering 
\includegraphics[angle=0,width=6cm]{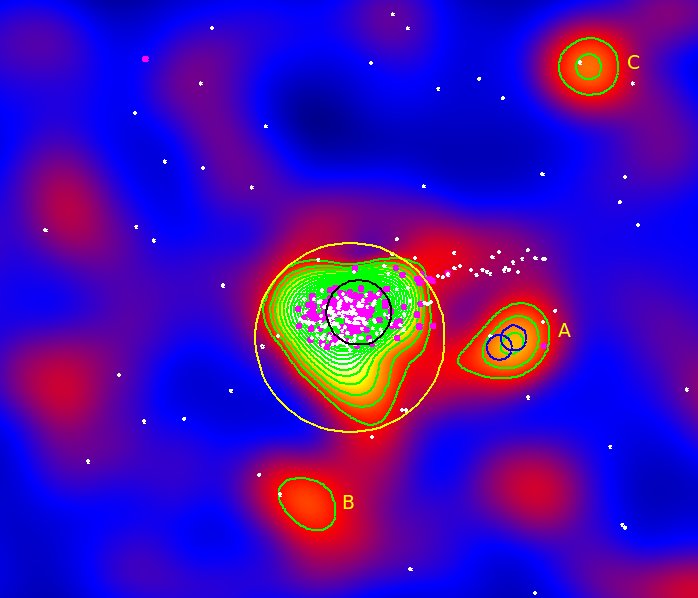}
%\mbox{\psfig{figure=A851_cont_ell_zspec.ps,width=7cm,clip=true}}
\caption{Same as Fig.~\ref{fig:cl0016} for Abell 851 (z=0.4069).
  The two medium-size blue circles show the two possible NED
  identifications of the western feature.
The magenta points correspond to the galaxies with spectroscopic redshifts
in the $0.385<z<0.425$ interval.}
\label{fig:a851}
\end{figure}

Abell~851 is a moderately massive ($M_{200}^{NFW}=5.5\times
10^{14}$~M$_\odot$ in M15) and highly substructured cluster (G14), and
the mass distribution derived from weak lensing (M15) is quite
similar to the galaxy density map shown in Fig.~\ref{fig:a851}. The
structure (labelled A) visible 15.0~arcmin (4.9~Mpc at the cluster
redshift) west of the cluster roughly corresponds to structure~5 in
M15 and is linked to Abell~851 by a filament that does not quite reach
a $3\sigma$ detection. It could be identified with two objects in NED:
either the galaxy group ABELL 0851:[KKN2011] West or the cluster WHL
J094203.2+465603 at photo$-z$=0.4606. A weak structure (B) is detected
18.8~arcmin (6.1~Mpc) south of the cluster, possibly linked to the
cluster, but the linking matter is not detected at $3\sigma$. A third
structure (C) is detected 32~arcmin (10.4~Mpc) north-west of the
cluster. Neither B nor C have identifications in NED. There are 208
spectroscopic redshifts in the $0.385<z<0.425$ range, mostly
concentrated in the cluster itself, and their histogram is quite
perturbed, as expected from a merging cluster.

\subsubsection{BMW-HRI J122657.3+333253 (z=0.8900)} %YES? 186.8 fig20

\begin{figure}[h!]
\centering 
\includegraphics[angle=0,width=6cm]{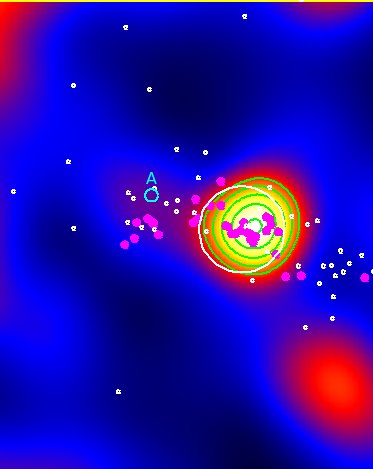}
%\mbox{\psfig{figure=BMW1226_b38coupe_cont_ell_zspec.ps,width=6cm,clip=true}}
\caption{Same as Fig.~\ref{fig:cl0016} for BMW-HRI J122657.3+333253
  (z=0.8900).  The magenta points correspond to the galaxies with
  spectroscopic redshifts in the $0.87<z<0.92$ interval.}
\label{fig:bmw1226}
\end{figure}

BMW-HRI J122657.3+333253 is the most distant cluster of the DAFT/FADA
sample and gave us the opportunity to discuss our method for
calculating significance contours, as described in Sect.~2.4.  Our
final choice was to cut the catalogue to eliminate the four sides of
the image where edge effects were present, as well as the zone that
was contaminated by NGC~4395. The result is shown in
Fig.~\ref{fig:bmw1226}.  Here, the cluster is detected at $7\sigma$,
but X-class~1808 is not detected (the cyan circle labelled A).  In
view of the differences in redshift, there is no physical link between
this cluster and BMW-HRI J122657.3+333253.

Thirty-four spectroscopic redshifts are available in the $0.87<z<0.92$
interval, out of which 24 are in the range $0.89<z<0.925$ .

\subsubsection{ [MJM98] 034 (z=0.383) } %YES 203.7 fig21

\begin{figure} 
\centering 
\includegraphics[angle=0,width=6cm]{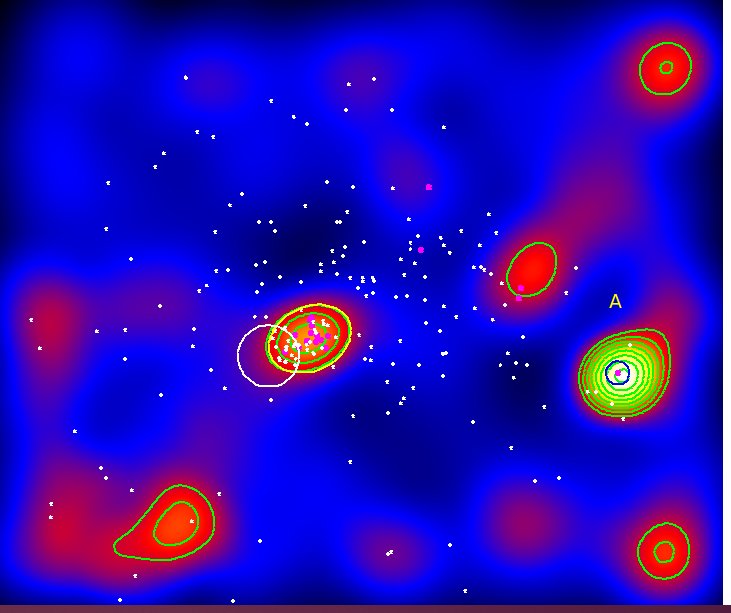}
%\mbox{\psfig{figure=MJM98_cont_ell_zspec2.ps,width=7cm,clip=true}}
\caption{Same as Fig.~\ref{fig:cl0016} for [MJM98] 034 (z=0.383).
The magenta points correspond to the galaxies with spectroscopic redshifts
in the $0.380<z<0.387$ interval.}
\label{fig:mjm98}
\end{figure}

[MJM98]~034 was discovered by McHardy et al. (1998) in a ROSAT image.
These authors considered its identification as certain, but gave an
uncertain redshift $z=0.595$, and due to the offset of 9.35~arcmin
relatively to the centre of the ROSAT field, the coordinates of this
cluster are probably not very accurate. The coordinates and redshift
published by McHardy et al. (1998) are those currently available in
NED. Based on XMM-Newton data, Mehrtens et al. (2012) did not give a
different or more precise position, and the redshift in their Table~3
is $z=0.60$. However, as briefly discussed by G14, the 16 galaxy
redshifts that we extracted from NED around [MJM98]~034 give a mean
value $z\sim 0.383$, therefore we chose to select cluster galaxies
assuming this value. In the density map of Fig.~\ref{fig:mjm98} the
cluster is detected at a $5\sigma$ level 4.5~arcmin north-west of its
NED position.  Of the 23 galaxies with spectroscopic redshifts in the
$0.380<z<0.387$ range found in the entire $1\times 1$~deg$^2$ field,
about 10 coincide with the structure that we identify as [MJM98]~34 in
the density map, suggesting that its identification with a cluster at
$z\sim 0.383$ is correct.  Several other structures are visible in
Fig.~\ref{fig:mjm98}, but no filament is detected between the various
structures at a 3$\sigma$ level. The bright feature (labelled A)
36.8~arcmin to the west can be identified with the cluster
WHL~J133249.0+374710 at photo$-z$=0.3821.  If [MJM98]~034 is indeed at
$z=0.383$, these two clusters could form a pair.  However, the
projected distance of these two clusters on the sky would be rather
large: about 11.5~Mpc.

We also tried to select cluster galaxies assuming $z\sim 0.6$ to place
the red sequence, but the detection level of the cluster is then below
$3\sigma$. This is another argument to assume that [MJM98]~34 is at
$z=0.383$.

\subsubsection{ 3C 295 (z=0.4600)} %YES 212.8 fig22

\begin{figure}[h!]
\centering 
\includegraphics[angle=0,width=6cm]{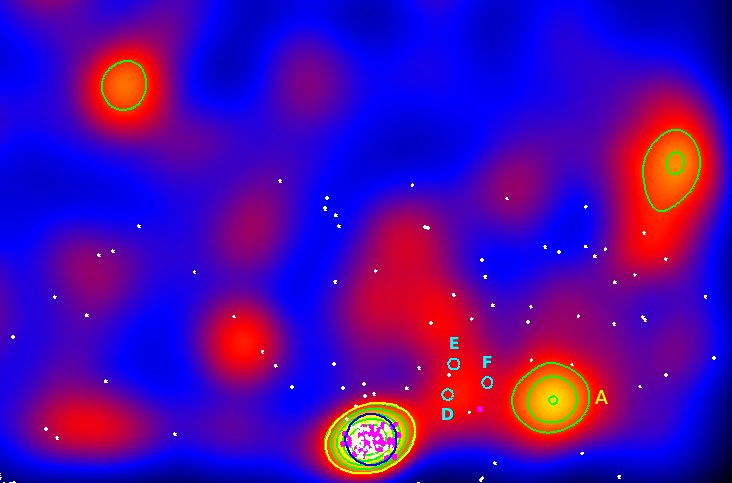}
%\mbox{\psfig{figure=3C295_cont_ell_zspec.ps,width=7cm,clip=true}}
\caption{Same as Fig.~\ref{fig:cl0016} for the 3C 295 cluster (z=0.4600).
The magenta points correspond to the galaxies with spectroscopic redshifts
in the $0.425<z<0.485$ interval.}
\label{fig:3c295}
\end{figure}

3C~295 shows a maximum extension of 3.6~Mpc. A second structure
(labelled A) is detected at a $5\sigma$ level west of the cluster but
has no identification in NED, and this is also the case for the two
other structures (B and C) farther away in the field.  The redshift
$z\sim 0.46$ was estimated by G14 and is different from the value
$z=0.2317$ given by NED.

Three clusters are found close to 3C~295 in NED (the small cyan
circles (D, E, and F) in Fig.~\ref{fig:3c295}) but are not detected in
our density map: W3-0995 of Durret et al. (2011b) at photo-$z$=0.75,
CFHT-W CL~J141045.3+521737 at photo-$z$=0.6854, and CFHT-W
CL~J141043.0+522038 at photo-$z$=0.4502.  There are 78 spectroscopic
redshifts in the $0.425<z<0.485$ range, concentrated in a circle of
1~Mpc radius, but quite spread out in redshift.

\subsubsection{RX J1524.6+0957 (z=0.5160)} %YES 231.1 fig23

\begin{figure} 
\centering 
\includegraphics[angle=0,width=6cm]{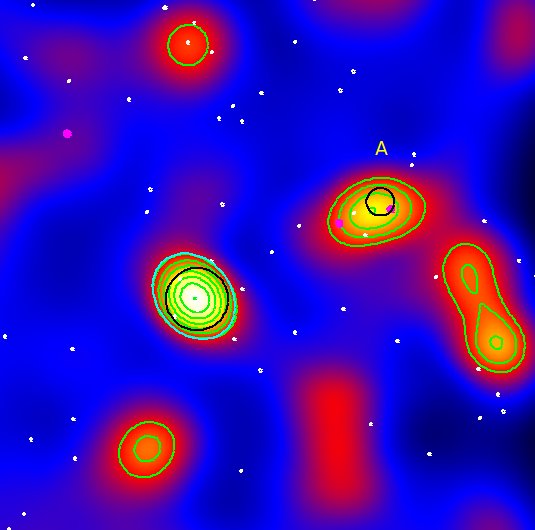}
%\mbox{\psfig{figure=RX1524_cont_ell_zspec.ps,width=7cm,clip=true}}
\caption{Same as Fig.~\ref{fig:cl0016} for RX J1524.6+0957 (z=0.5160).
The magenta points correspond to the galaxies with spectroscopic redshifts
in the $0.505<z<0.520$ interval.}
\label{fig:rx1524}
\end{figure}

Figure~\ref{fig:rx1524} shows a quite massive cluster. A
second rather bright structure (labelled A) is detected 17.9~arcmin
north-west of RX~J1524.6+0957 and can probably be identified with the
cluster GMBCG~J230.90400+10.10080 at photo$-z=0.517$, in which case
the two clusters could form a pair at a projected distance of about
6.7~Mpc. However, we do not detect matter between these two clusters.
Several other structures are detected in the field but have no NED
identification, and only three spectroscopic redshifts are available
in the $0.505<z<0.520$ range.

\subsubsection{OC02 J1701+6412 (z=0.4530)} %YES 255.3 fig24

\begin{figure}[h!]
\centering 
\includegraphics[angle=0,width=6cm]{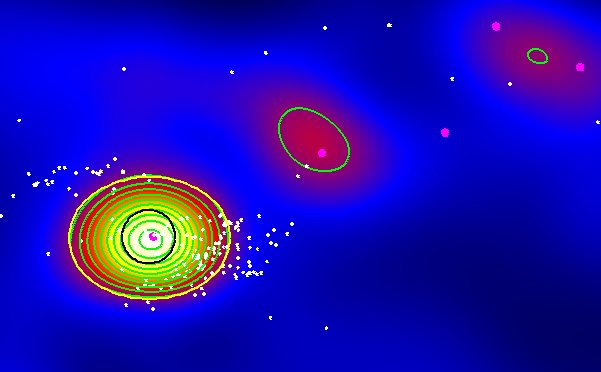}
%\mbox{\psfig{figure=OC02gi_cont_ell_zspec.ps,width=7cm,clip=true}}
\caption{Same as Fig.~\ref{fig:cl0016} for OC02 J1701+6412 (z=0.4530).
The magenta points correspond to the galaxies with spectroscopic redshifts
in the $0.44<z<0.46$ interval.}
\label{fig:oc02}
\end{figure}

OC02~J1701+6412 is not a very massive cluster
($M_{200}^{NFW}=3.6\times 10^{14}$~M$_\odot$ in M15) and shows no
obvious substructure in Fig.~\ref{fig:oc02}. M15 found a double
structure along a roughly north--south direction, but their structure 3
is most probably not at the cluster redshift, since we do not detect
it here.  Neither do we detect the M15 structure 2, which is
expected because it is identified with a lower redshift cluster that
we did not expect to detect here. We detect at $3\sigma$ two small
structures farther out to the north-west, but they have no
identification in NED and our detections are not significant enough to
conclude that OC02 belongs to a supercluster. Only five spectroscopic
redshifts are available in the range $0.450\leq z \leq 0.458$.

\subsubsection{RX J2328.8+1453 (z=0.4970)} %YES 352.2 fig25

\begin{figure} 
\centering 
\includegraphics[angle=0,width=6cm]{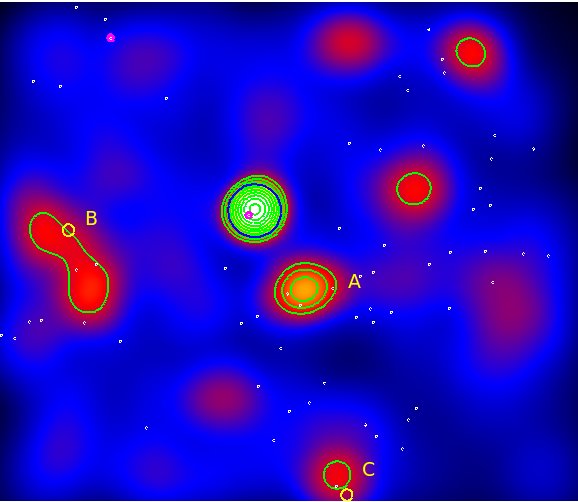}
%\mbox{\psfig{figure=RX2328gi_cont_ell_zspec.ps,width=7cm,clip=true}}
\caption{Same as Fig.~\ref{fig:cl0016} for RX J2328.8+1453 (z=0.4970).
The magenta points correspond to the galaxies with spectroscopic redshifts
in the $0.48<z<0.50$ interval.}
\label{fig:rx2328}
\end{figure}

RX~J2328.8+1453 appears to be very bright on the density map
(Fig.~\ref{fig:rx2328}) and is a moderately massive cluster
($M_{200}^{NFW}=5.4\times 10^{14}$~M$_\odot$ in M15). The structure
labelled A corresponds to the Pegasus dwarf galaxy, which is at much
lower redshift, but so bright that it still somewhat contaminates
the density map. There are two other structures, one (labelled B)
19.4~arcmin east and one (C) 31.0~arcmin south of the cluster that can
be identified in NED with the clusters GMBCG
J352.53003+14.85291 at $z=0.478$ and GMBCG J352.04865+14.39520 at
$z=0.416,  $ respectively. If we consider that RX~J2328.8+1453 and
GMBCG~J352.53003+14.85291 form a pair, their projected distance would
be about 7.1~Mpc.  Two other weak structures are detected at $3\sigma$
in the density map, but no filament system is observed. Only two
spectroscopic redshifts are available at $z=0.4985$ and 0.49.

\subsection{Three clusters without significantly detected 
neighbouring structures or filaments}

The following clusters are clearly detected on the density maps,
but show no significantly detected neighbouring structures or
filaments.

\subsubsection{MACS J0454.1-0300 (z=0.5377)} %YES 73.5 fig26

\begin{figure} 
\centering 
\includegraphics[angle=0,width=6cm]{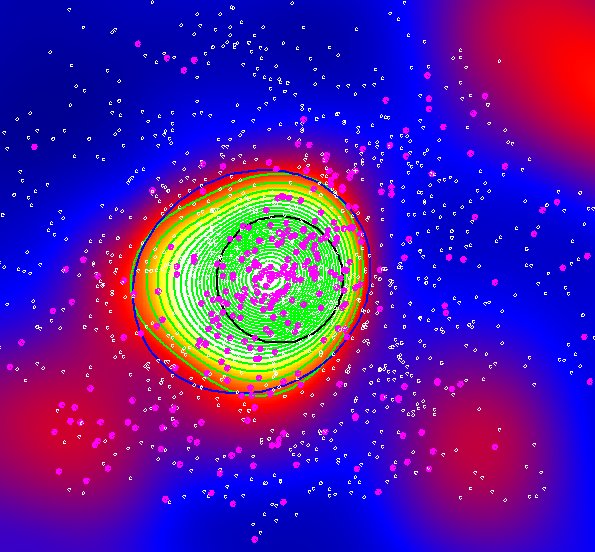}
%\mbox{\psfig{figure=MACS0454gZ_cont_ell_zspec.ps,width=7cm,clip=true}}
\caption{Same as Fig.~\ref{fig:cl0016} for MACS J0454.1-0300 (z=0.5377).
The magenta points correspond to the galaxies with spectroscopic redshifts
in the $0.522<z<0.553$ interval.}
\label{fig:macs0454}
\end{figure}

The galaxy density map (Fig.~\ref{fig:macs0454}) shows no obvious
elongation or deformation, although the weak-lensing analysis detects
two peaks (M15). However, we can note that the peak of the density map
(where the contours reach a maximum) does not coincide with the
position of the cluster given by NED, the displacement being about
0.011~deg, corresponding to 0.25~Mpc at the cluster distance.
According to the strong-lensing analysis made by Zitrin et al. (2011),
MACS~J0454.1-0300 is a rather low-mass cluster, with a M$_{500}$ mass
of the order of only $4\times 10^{13}$~M$_\odot$. No other structure
is detected in the density map, therefore we only show a zoom on the
cluster.  There are 343 spectroscopic redshifts in the $0.522<z<0.553$
range.

\subsubsection{MS 2053.7-0449 (z=0.5830) and 
CXOSEXSI J205617.1-044155 (z=0.6002)} %YES 314.1 fig27

\begin{figure}[h!]
\centering 
\includegraphics[angle=0,width=6cm]{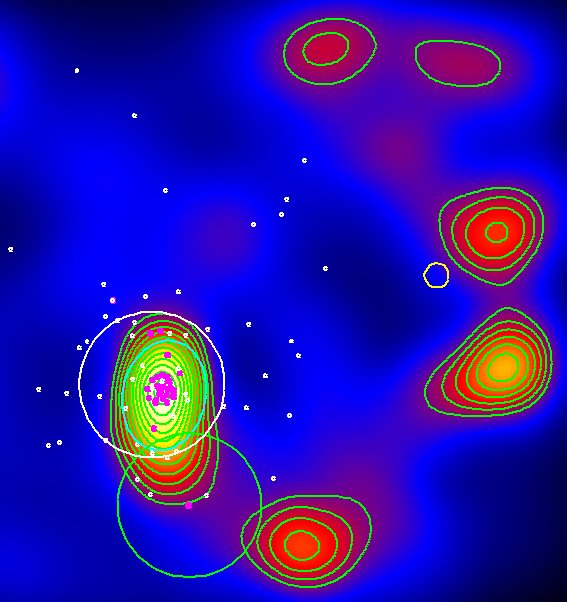}
%\mbox{\psfig{figure=MS2053VZ_cont_ell_zspec.ps,width=7cm,clip=true}}
\caption{Same as Fig.~\ref{fig:cl0016} for MS 2053.7-0449 in white and
  CXOSEXSI J205617.1-044155 in cyan (z=0.5830 and 0.6002,
  respectively). The magenta points correspond to the galaxies with
  spectroscopic redshifts in the $0.565<z<0.610$ interval.}
\label{fig:ms2053}
\end{figure}

MS 2053.7-0449 is a massive cluster ($M_{200}^{NFW}=1.07\times
10^{15}$~M$_\odot$ in M15) detected at the $5\sigma$ level, but its
companion cluster CXOSEXSI~J205617.1-044155 is only visible as a faint
extension to the south of MS~2053.7-0449 on the density map (see
Fig.~\ref{fig:ms2053}). Several other structures are detected, but the
only galaxy cluster with a redshift available in NED is [BGV2006]~078
at z=0.277 (the yellow circle); it is not detected here, as expected
in view of its redshift. None of the other structures are found in
NED, so we have no unambiguous information on the environment of these
two clusters. There are 28 spectroscopic redshifts in the
$0.565<z<0.610$ range, mostly concentrated in MS~2053.7-0449.

\subsection{Four clusters whose contrast is too low to be detected at
  $3\sigma$ on the density maps}

Four clusters at redshifts between $z=0.66$ and $z=0.87$ are of too
low contrast to be clearly visible on our density maps, at least at
the positions given by NED. We briefly give a few indications on these
clusters below.

The two clusters SEXCLAS~12 and SEXCLAS~13 do not appear clearly on
our density map. Since the exposure times of our $r'$ and $z'$ images
are 2040 and 10800~seconds, respectively, we ought to detect them if
they were at their NED redshifts and had masses higher than $2\times
10^{14}$~M$_\odot$ (the minimum mass of the DAFT/FADA survey clusters,
as derived from X-ray ROSAT data).  This means that if these clusters
are not detected on the density map, it is because they are more
distant and/or less massive than expected. We checked if they could be
at higher redshifts than the NED values.  SEXCLAS~12 and 13 were
discovered by Kolokotronis et al. (2006), who estimated their
respective photometric redshifts to be $z=0.61$ and 0.58. Among the
spectroscopic redshifts that we gathered from NED, we found two
galaxies located very close to the centres of these two clusters and
that are probably the BCGs.  At 19~arcsec of the centre of SEXCLAS~12
a galaxy is located at RA=163.15542$^\circ$and DEC=57.51778$^\circ$
with $z=0.708$, and at 42~arcsec of the centre of SEXCLAS~13 a galaxy
is located at RA=163.21459$^\circ$and DEC=57.53361$^\circ$, with
$z=0.664$.  Therefore we believe that the correct redshifts of
SEXCLAS~12 and SEXCLAS~13 are $z=0.708$ and $z=0.664,$
respectively. The histogram of the redshifts we were able to gather is
quite flat , which makes it impossible to derive dynamical properties
for this system. Since these redshifts are not extremely high, the
fact that we do not detect SEXCLAS~12 and SEXCLAS~13 on the density
map suggests that these are not massive clusters, in agreement with
their non-detection by XMM-Newton (see G14, Appendix C.5.) and by
their faintness that impeded deriving their galaxy luminosity function
(Martinet et al. 2015a). These two clusters should probably not have
been included in the DAFT/FADA cluster survey.

The cluster RCS~J1620.2+2929 ($z=0.87$) is not detected on the density
map either, probably because the exposure time in the $i'$ filter is
too short (516~s) to detect such a distant object. Only the cluster
redshift given by NED is available in the $0.85<z<0.89$ range.

For NEP~200 (z=0.6909), the position given in Table~1 coincides with
the BCG, but we detect the cluster on the density map only at the
$2.3\sigma$ level. This is rather surprising since M15 give a high
mass for this cluster, if with a large error bar
($M_{200}^{NFW}=(14.8\pm 6.1)\times 10^{14}$~M$_\odot$). As for
BMW-HRI J122657.3+333253, we tried to cut the density map and
eliminate a very bright feature at the western edge of the image, but
this only improves the detection level to $2.9\sigma$. The new
redshifts measured with GTC are given in Table~\ref{tab:GTCNEP}, and
the redshift histogram including these new redshifts and those found
in NED is shown in Fig.~\ref{fig:histozNEP200}.

We can therefore conclude that clusters with redshifts of the order of
$z\sim 0.66$ or higher will only be detected on density maps such as those that
we computed if they are massive (with the obvious condition that the
exposure times of the images are sufficient).

\section{Discussion and conclusions}

The DAFT/FADA survey, started several years ago (principal
investigators C.~Adami, D.~Clowe, and M.~Ulmer), comprises a sample of
90 massive clusters (mass higher than $2\times 10^{14}$~M$_\odot$,
derived from ROSAT X-ray data) in the medium-high redshift range
$0.4\leq z\leq 0.9$ and with HST data available. For a large part
of them, we gathered deep ground-based images in several optical bands
and in one infrared band. To analyse the large-scale environment of these
clusters, we here selected the thirty clusters
for which large field images obtained with CFHT/Megacam or
Subaru/SuprimeCam were available (either from our own data or from the
archives).  We limited our analysis to the clusters with at least two
bands obtained with the same instrument except for two clusters, and
bracketing whenever possible the 4000~\AA\ break. In this way, our
sample included thirty clusters: nineteen with CFHT/Megacam data, nine
with Subaru/SuprimeCam data, and two with mixed CFHT/Megacam and
Subaru/SuprimeCam data.

For each cluster, we selected galaxies that had a high probability of
belonging to the cluster based on colour-magnitude diagrams. The
influence of the position of the red sequence in the colour-magnitude
diagram and of the width of the galaxies selected on either side of
the red sequence were discussed in Sect.~2.3.  and are illustrated in
Appendix~A. 

Based on these catalogues of cluster galaxies, we then computed galaxy
density maps by applying an adaptive kernel method. This technique is
well suited here, since it allows detecting weak structures, but at
the same time gives more precise results in the dense zones where the
signal is high (i.e. where there are many cluster galaxies). We then
estimated for each density map the value of the background and drew
contours at $3\sigma$ and higher above this value.  We also computed a
mean background for a given telescope and set of filters to minimize
the effect of cosmic variance (the background is not exactly the same
for all the clusters).  In most cases, the clusters are detected at
the same significance level as with the local background.  However,
the background in the density map of [MJM98]~34 is notably lower than
for the other clusters, which means that if we compute significance
levels above the average value, we no longer detect [MJM98]~34 at
$3\sigma$.  We therefore kept our original analysis (with a local
background), but we recall that computing the local background around
each cluster can be a source of inhomogeneity among clusters.

We presented and discussed the results individually for each cluster
in Sect.~3. We discuss the influence of bright sources on computing
the significance levels of the features that we detect in Appendix B.
We find clear elongations in twelve clusters out of thirty, with sizes
that can reach up to 7.6~Mpc (see Sect. 3.1. and Table~2). The
distinction between an elongation and a filament is not obvious. We
tried to quantify here the maximum sizes of the $3\sigma$ detection
contours, and in most cases we only detected asymmetries in the
cluster shapes.  In some cases such as ZwCl~1332.8+5043, there are
only three small elongations and the $3\sigma$ detection contour is
almost circular, but we included it among the twelve clusters with
elongations because of its size of $5.8\times 5.4$~Mpc$^2$, which is
relatively large compared to a typical cluster size.

Of the twelve clusters that are clearly elongated, ten have redshifts
$z<0.6$, one is at $z=0.690$ (MACS~J0744.9+3927), and only one is
really distant (RX~J1716.4+6708 at $z=0.8130$). Therefore, we see that
weak extensions or filamentary structures are easier to detect for
relatively nearby clusters than for more distant ones, as
expected. The masses of six of these twelve clusters were computed
based on the weak-lensing analysis by M15, and they are all in the
medium-high to high mass range, between $(6.4\pm 3.1)\times
10^{14}$~M$_\odot$ and $(22.6\pm 6.2)\times 10^{14}$~M$_\odot$. The
fact that extensions are easier to detect at low redshifts is
confirmed by the fact that the clusters whose contrast is too low to
be unambiguously detected on our density maps and observed with long
exposure times (see Sect. 3.4) are all at relatively high redshifts
(in the range $z>0.66$).

Eleven other clusters have neighbouring structures, but the zones
linking these structures to the cluster are not detected at the
3$\sigma$ level in the density maps, and three clusters show no
extended structure and no neighbours.  The characteristics of these
fourteen clusters are less clear than those described above, since
they span the entire ranges of redshifts and masses of our sample.

We are aware that separating clusters between Sects. 3.1 and 3.2 may
be somewhat arbitrary. Since the red sequences are computed with
various colours, slightly different galaxy populations are selected
depending on the available data, and since different galaxy
populations are clustered in different way, our analysis is not fully
homogeneous. Unfortunately, we have no cluster for which both Megacam
and Subaru data are available in all the bands, so we cannot test how
the red sequence changes with all the colour selections. Our
classification gives indications on what is detected around the
clusters of our sample, however.

The number of redshifts available in the cluster range varies strongly
from one cluster to another (between one and several hundred).  We
detect filaments or extensions in clusters both with many and few
available spectroscopic redshifts, which tends to confirm that our
selection of cluster galaxies based on colour-magnitude diagrams is
reliable. This was also shown to be the case by \kart.  One important
consequence of the fact that we did not detect strong filaments (except
for MACS~J0717+3745) is that the clusters that we
analysed between redshifts $z=0.4$ and $z=0.9$ have already formed,
and even if some of them show evidence that they are undergoing a
merger, they are probably in the final stages of their formation
process and do no longer accrete a large quantity of matter
through filaments.

We can note that most of the clusters that were only weakly detected
or even undetected by our method have very few spectroscopic
redshifts, which clearly shows that redshifts help to define the
cluster red sequence more accurately and thus to obtain a better
selection of the cluster galaxies.  For the clusters of our sample
with several hundred available spectroscopic redshifts in the cluster
redshift range, we are beginning a detailed dynamical analysis to
determine whether the cluster properties seen on the density maps
(such as their relaxed or perturbed aspect) correlate with their
dynamical properties. This will also allow a mass estimate independent
of that obtained either from X-rays (G14) or weak lensing (M15).

The search for extensions and filaments around clusters is of
cosmological importance since they allow tracing the directions along
which clusters are still accreting galaxies, groups, or small clusters,
and in this way allows better understanding the formation and evolution of
clusters. The exact definition of filaments is still under debate, as was
discussed for example by Pimbblet (2005). It is obviously linked both
to the geometry and to the density of the structures, and detecting
more filaments should allow a more accurate definition of what
filaments are.  With this question in mind, we are in the process of
searching for filaments around the thousands of candidate clusters
detected in the CFHTLS (Durret et al. 2011b and references therein)
and SDSS Stripe~82 (Durret et al. 2015) surveys, with the aim to be able
to assemble statistics on filaments around clusters (Sarron et al. in
preparation). This large search will be possible once the
  procedure is made 100\% automatic.  

Although we did not address this question here, it would be interesting
to conduct a total census of the mass contained in filaments, that is, the
total baryonic mass, which consists of an intergalactic medium plus galaxies,
combined with the non-baryonic cold dark matter mass. Eckert et
al. (2015) have recently detected filaments both in the optical and
X-rays around the Pandora cluster Abell~2744 and concluded that the
properties of these filaments ``supported the picture in which a large
fraction of the Universe's baryons are located in the filaments of the
cosmic web''.  Future satellite missions that will use weak lensing such
as EUCLID, W-First, together with a relatively wide field of view and
high sensitivity X-ray mission such as Athena, will be able to provide
the necessary data to be compared with ongoing $\Lambda$CDM-based
simulations of the cosmic web (e.g Haider et al, 2015).

\begin{acknowledgements}

  We thank the referee for the interesting comments.  We are
  grateful to Emmanuel Bertin for many enlightening discussions. FD
  acknowledges long-term support from CNES. IM acknowledges financial
  support from the Spanish Ministry of Economy and Competitiveness
  through the grant AYA2013-42227P.

\end{acknowledgements}

\appendix

\section{Testing the chosen parameters of the red sequence 
that were used to build density maps}

\begin{figure}[h!] %figA1
\centering 
\includegraphics[angle=0,width=6cm]{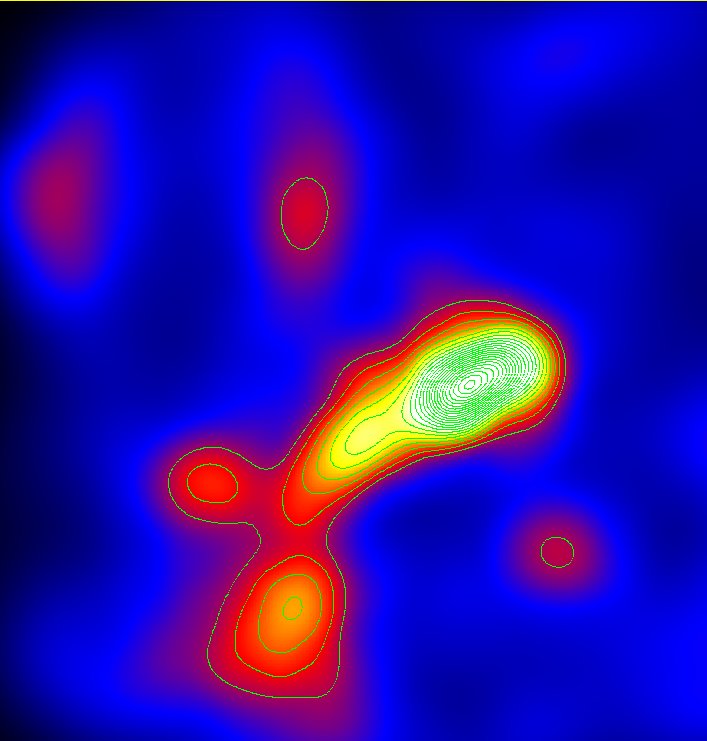}
\includegraphics[angle=0,width=6cm]{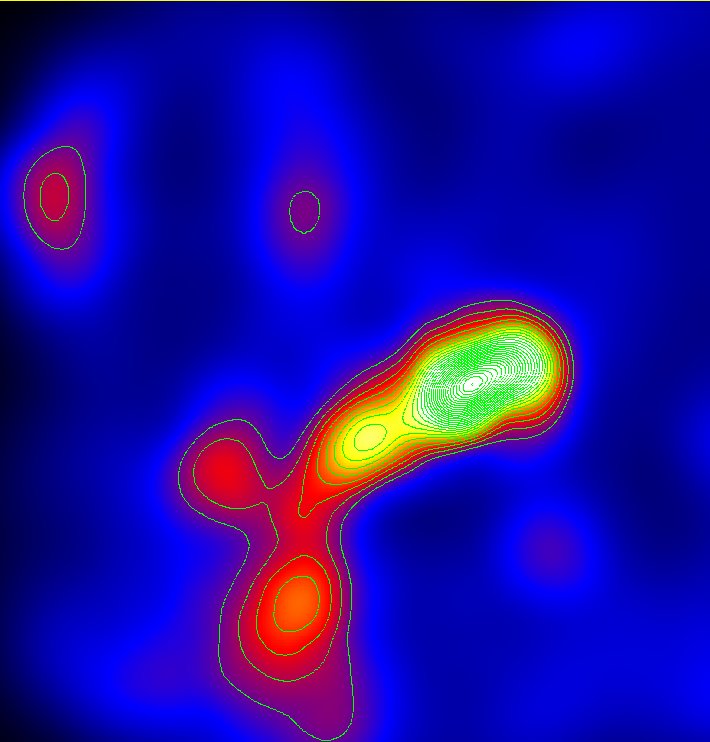}
\includegraphics[angle=0,width=6cm]{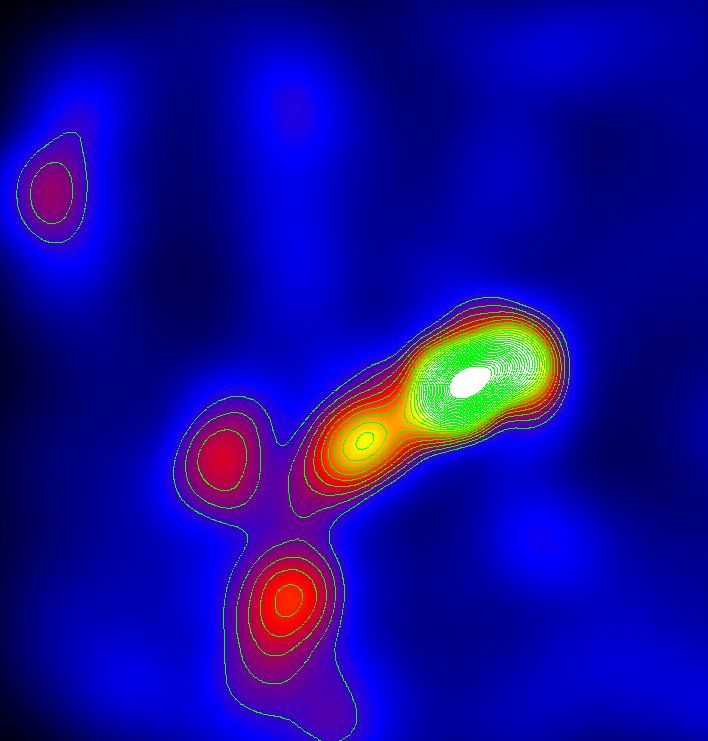}
\includegraphics[angle=0,width=6cm]{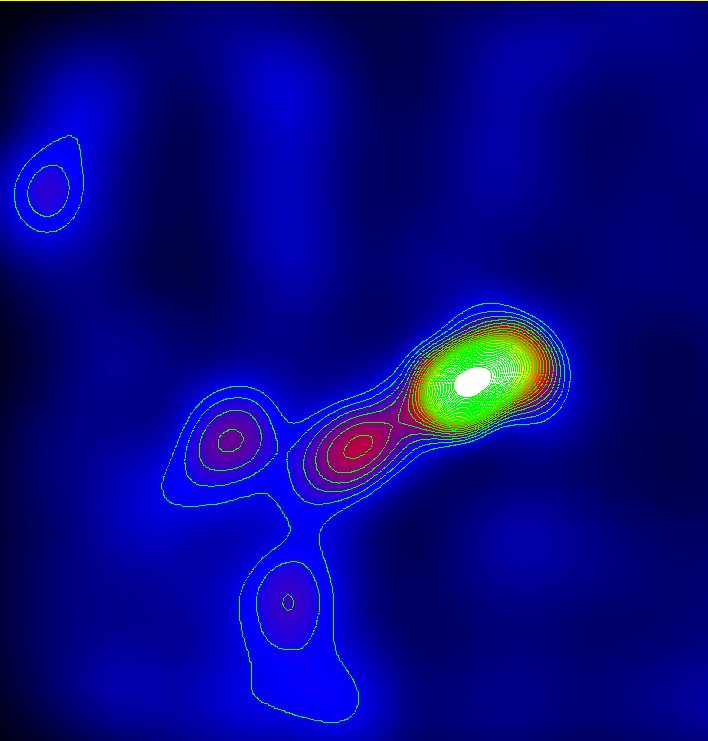}
%\mbox{\psfig{figure=MACS0717_mean_dens_b250d03_cont.ps,width=6cm,height=6cm}}
%\mbox{\psfig{figure=MACS0717_mean_dens_b275d03_cont.ps,width=6cm,height=6cm}}
%\mbox{\psfig{figure=MACS0717_mean_dens_b300d03_cont.ps,width=6cm,height=6cm}}
%\mbox{\psfig{figure=MACS0717_mean_dens_b325d03_cont.ps,width=6cm,height=6cm}}
\caption{Density maps obtained for different heights of the red
  sequence $V-I=-0.0436\times I+b$ (see Sect.~2.2). From top to
  bottom: $b$=2.50, 2.75, 3.00, and 3.25. Galaxies within $\pm 0.3$
  magnitudes from the red sequence are selected. The dynamic range is
  the same in all the figures, but the contour levels are estimated
  for each image as described in Sect.~2.2.}
\label{fig:0717_varyb}
\end{figure}

\begin{figure}[h!] %figA2
\centering 
\includegraphics[angle=0,width=6cm]{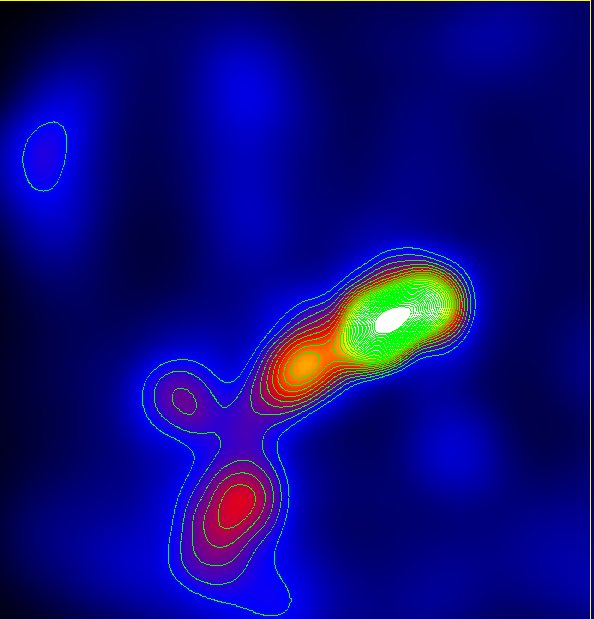}
\includegraphics[angle=0,width=6cm]{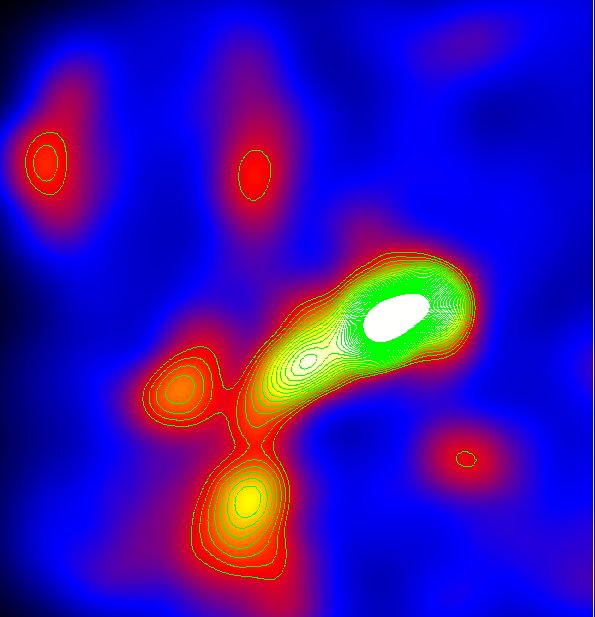}
%\mbox{\psfig{figure=MACS0717_mean_dens_b275d02_cont.ps,width=6cm,height=6cm}}
%\mbox{\psfig{figure=MACS0717_mean_dens_b275d04_cont.ps,width=6cm,height=6cm}}
\caption{Same as Fig.~\ref{fig:0717_varyb} with $b=2.75$, but
  selecting galaxies within $\pm 0.2$ (top) and $\pm 0.4$ (bottom) of
  the red sequence.}
\label{fig:0717_varyd}
\end{figure}

We illustrate here the influence that the selection
of the galaxies along the red sequence has on the density maps. Figure~\ref{fig:0717_varyb}
shows the influence of the height of the red sequence, and
Fig.~\ref{fig:0717_varyd} shows how the density maps change if the
width around the red sequence is modified. Details are given in Sect.~2.3.

\section{Testing the determination of the cluster detection levels}

\begin{figure}[h!] %fig B1
\centering 
\includegraphics[angle=0,width=6cm]{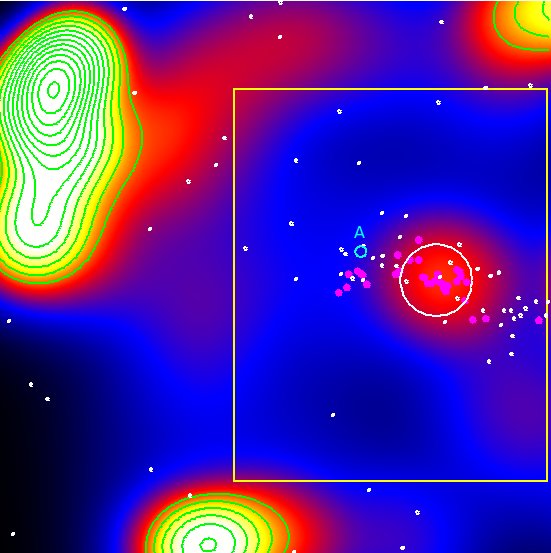}
\includegraphics[angle=0,width=6cm]{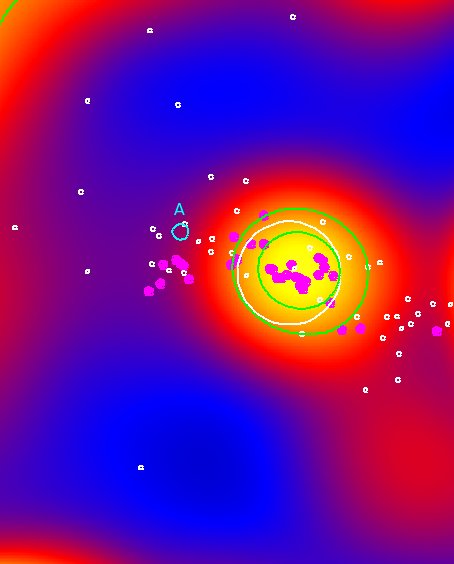}
\includegraphics[angle=0,width=6cm]{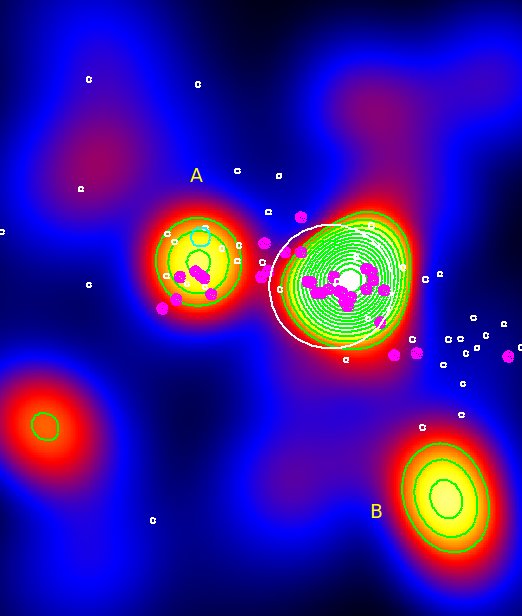}
%\mbox{\psfig{figure=BMW1226_b38_cont_ell_zspec.ps,width=6cm,clip=true}}
%\mbox{\psfig{figure=BMW1226_b38sm_cont_ell_zspec.ps,width=6cm,clip=true}}
%\mbox{\psfig{figure=BMW1226_b38sm2_cont_ell_zspec.ps,width=6cm,clip=true}}
\caption{Same as Fig.~\ref{fig:cl0016} for BMW-HRI J122657.3+333253
  (z=0.8900). The top figure shows the density map for the entire
  Subaru/SuprimeCam field with the yellow rectangle indicating
    the subregion considered in the two figures just below. The
  middle figure shows the subimage extracted (within the yellow
    rectangle) from the initial density map with contour levels
    recalculated, based on the background of this new subimage. The
    bottom image was computed after extracting from the initial
    catalogue of cluster galaxies the galaxies belonging to the yellow
    rectangle and recomputing the density map. See text for more
  explanations.  The magenta points correspond to the galaxies with
  spectroscopic redshifts in the $0.87<z<0.92$ interval.}
\label{fig:bmw1226tests}
\end{figure}

We illustrate here the influence of bright structures on the
determination of the background level of the density maps that has
important consequences on the significance levels of the detections,
based on the extreme example of BMW-HRI J122657.3+333253.  See Sect.
2.4 for details.

\section{New spectroscopic redshifts obtained with the GTC
and various redshift histograms}

\begin{table}[t!]
  \caption{New spectroscopic redshifts obtained with GTC
    in the field of ZwCl~1332.8+5043. The columns are (1)~running number, 
    (2)~RA (2000.0), (3)~DEC (J2000.0), (4)~spectroscopic redshift, 
    (5)~flag (see text), (6)~Megacam $r'$ band magnitude. }
\begin{tabular}{rrrrrr}
\hline
\hline
         &           &           &   &      & \\
Running  & RA        & DEC       & z~~ & flag & $r'$ band \\
number   & (J2000.0) & (J2000.0) &   &      & magnitude \\
\hline
   &           &          &        &   & \\
 1 & 203.54648 & 50.54227 & 0.2972 & 3 & 20.455 \\
 2 & 203.56148 & 50.55794 & 0.5505 & 3 & 21.135 \\
 3 & 203.56590 & 50.46159 & 0.4047 & 2 & 21.261 \\
 4 & 203.57019 & 50.56184 & 0.2964 & 3 & 22.256 \\
 5 & 203.57242 & 50.53869 & 0.6155 & 3 & 22.302 \\
 6 & 203.57759 & 50.55912 & 0.6316 & 3 & 21.407 \\
 7 & 203.57977 & 50.43289 & 0.3317 & 2 & 21.375 \\
 8 & 203.58572 & 50.51764 & 0.5854 & 9 & 20.470 \\
 9 & 203.58656 & 50.43459 & 0.2673 & 2 & 22.384 \\
10 & 203.58722 & 50.52655 & 0.6277 & 4 & 21.594 \\
11 & 203.59153 & 50.51581 & 0.6097 & 3 & 21.878 \\
12 & 203.59155 & 50.51013 & 0.1486 & 9 & 20.067 \\
13 & 203.59267 & 50.50076 & 0.2690 & 2 & 21.384 \\
14 & 203.59357 & 50.51333 & 0.6100 & 4 & 20.749 \\
15 & 203.59570 & 50.50032 & 0.2617 & 2 & 20.315 \\
16 & 203.60213 & 50.53054 & 0.6208 & 3 & 22.258 \\
17 & 203.60523 & 50.50613 & 0.4438 & 2 & 19.105 \\
18 & 203.61669 & 50.50579 & 0.6162 & 4 & 21.184 \\
19 & 203.61814 & 50.47468 & 0.2623 & 3 & 19.115 \\
20 & 203.62027 & 50.55429 & 0.2770 & 3 & 19.227 \\
21 & 203.62365 & 50.48701 & 0.4425 & 3 & 20.830 \\
22 & 203.62508 & 50.45514 & 0.0831 & 2 & 15.874 \\
23 & 203.62663 & 50.44983 & 0.0842 & 2 & 17.348 \\
24 & 203.62823 & 50.47422 & 0.2662 & 4 & 20.156 \\
25 & 203.63465 & 50.47806 & 0.0838 & 3 & 20.040 \\
26 & 203.63623 & 50.46817 & 0.0882 & 2 & 19.239 \\
27 & 203.63952 & 50.46948 & 0.0844 & 3 & 20.192 \\
28 & 203.64069 & 50.48123 & 0.2661 & 2 & 19.640 \\
29 & 203.65446 & 50.48689 & 0.0845 & 3 & 17.292 \\
30 & 203.65843 & 50.52982 & 0.3342 & 3 & 21.391 \\
   &           &          &        &   & \\
\hline
\end{tabular}
\label{tab:GTCZw}
\end{table}

\begin{table}[t!]
  \caption{New spectroscopic redshifts obtained with GTC
    in the field of NEP~200. The columns are (1)~running number, 
    (2)~RA (2000.0), (3)~DEC (J2000.0), (4)~spectroscopic redshift, 
    (5)~flag (see text), (6)~Megacam $r'$ band magnitude. }
\begin{tabular}{rrrrrr}
\hline
\hline
         &           &           &   &      & \\
Running  & RA        & DEC       & z~~ & flag & $r'$ band \\
number   & (J2000.0) & (J2000.0) &   &      & magnitude \\
\hline
   &           &          &        &    & \\
 1 & 269.26295 & 66.49133 & 0.0000 & 4 & 20.46 \\
 2 & 269.26708 & 66.47758 & 0.6105 & 2 & 20.87 \\
 3 & 269.27117 & 66.54886 & 0.4156 & 3 & 21.28 \\
 4 & 269.27431 & 66.52642 & 0.1749 & 4 & 23.50 \\
 5 & 269.27518 & 66.55717 & 0.0000 & 2 & 21.48 \\
 6 & 269.30138 & 66.46396 & 0.2624 & 3 & 18.10 \\
 7 & 269.30734 & 66.48506 & 0.0000 & 2 & 18.86 \\
 8 & 269.30746 & 66.45269 & 0.7771 & 9 & 21.98 \\
 9 & 269.31076 & 66.46900 & 0.5625 & 2 & 21.38 \\
10 & 269.31186 & 66.49042 & 0.6110 & 2 & 21.35 \\
11 & 269.31965 & 66.44800 & 0.2597 & 2 & 19.99 \\
12 & 269.32835 & 66.52599 & 0.5434 & 3 & 21.17 \\
13 & 269.33149 & 66.52576 & 0.6936 & 4 & 19.97 \\
14 & 269.33663 & 66.51119 & 0.4632 & 4 & 21.84 \\
15 & 269.34054 & 66.48596 & 0.5916 & 2 & 21.04 \\
16 & 269.34938 & 66.57275 & 0.0560 & 4 & 18.47 \\
17 & 269.35608 & 66.54448 & 0.6938 & 2 & 20.80 \\
18 & 269.35675 & 66.49611 & 0.4240 & 2 & 20.72 \\
19 & 269.35737 & 66.51684 & 0.0000 & 3 & 21.62 \\
20 & 269.37533 & 66.50135 & 0.3077 & 4 & 20.12 \\
21 & 269.38035 & 66.56889 & 0.6102 & 2 & 21.33 \\
22 & 269.38913 & 66.57149 & 0.7490 & 2 & 21.74 \\
23 & 269.39112 & 66.51077 & 0.0000 & 4 & 21.32 \\
24 & 269.41683 & 66.54119 & 0.0000 & 4 & 20.82 \\
   &           &          &        &   & \\
\hline
\end{tabular}
\label{tab:GTCNEP}
\end{table}

A number of spectroscopic redshifts were obtained through multi-slit
spectroscopic data obtained with OSIRIS/MOS on the Gran Telescopio
Canarias (GTC) on May 22,$^{}$ 2014, for the clusters
ZwCl~1332.8+5043 and NEP~200. Images previously obtained with
OSIRIS/GTC in the imaging mode were used to make the masks. Two masks
were made for each cluster, allowing slit spectroscopy of about
25 objects per mask. The grism R300R was used, which provides a
dispersion of 7.74 \AA/pixel in the wavelength range from 4800 to
10000 \AA. A total of 3000 seconds were exposed for each
mask. Arc-lamp exposures were taken for wavelength calibration.  The
seeing FWHM was measured in the acquisition images and reached
from 0.9 to 1.2~arcsec. The observing conditions were mostly
photometric. The standard star GD153 was observed for flux
calibration.

For the data reduction we applied the IRAF-based GTCMOS pipeline
(http://www.inaoep.mx/~ydm/gtcmos/gtcmos.html). We applied the usual
steps for bias subtraction and illumination correction, wavelength
calibration with a combination of Xe, Ne and HgAr lamp exposures, 
sky subtraction, and 1D spectrum extraction.

The spectroscopic redshifts were measured by cross-correlating them
with templates with the EZ code (Garilli et al. 2010).  A flag between
1 (worst) and 4 (best) was given to each measurement. Flag 1 means
that we have a 50$\%$ chance to have the correct redshift estimate,
flag~2 indicates a 75$\% $ chance, flag~3 a chance of 95$\%$, and flag~4 a chance higher than 99$\%$. Redshift
measurements with flag=1 were discarded.

We list in Tables~\ref{tab:GTCZw} and \ref{tab:GTCNEP} the values of
the new redshifts measured with GTC, together with the $r'$ band
magnitudes measured in our Megacam images.

We show below the galaxy spectroscopic redshift histograms zoomed on
the cluster redshift range for these two clusters, as well as for
three other clusters for which redshift histograms were not previously
given in our data paper (G14).  These histograms include all the NED
data and also the GTC data for Zw~1332 and NEP~200.

\begin{figure}[h!] %figC1
\centering 
\includegraphics[angle=0,width=6cm]{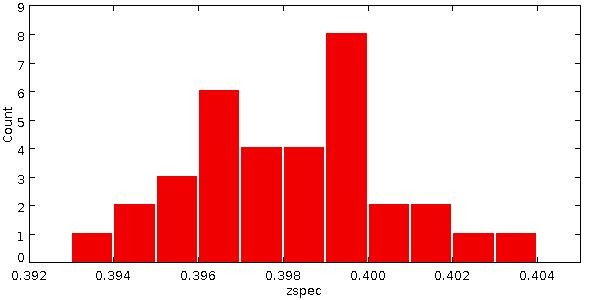}
%\mbox{\psfig{figure=PDCS18_histoz.ps,width=7cm}} %36.85
\caption{Galaxy spectroscopic redshift histogram in the field of
  PDCS~018.}
\label{fig:histozPDCS18}
\end{figure}

\begin{figure}[h!] %fig C2
\centering 
\includegraphics[angle=0,width=6cm]{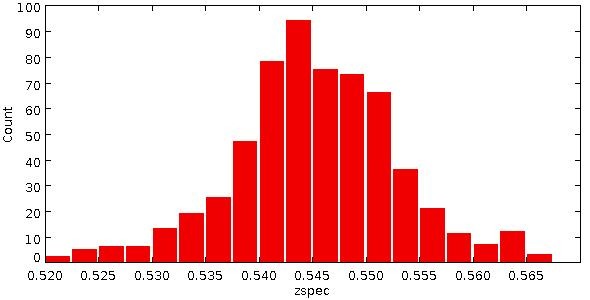}
%\mbox{\psfig{figure=MACS0717_histoz.ps,width=7cm}} %109.39
\caption{Galaxy spectroscopic redshift histogram in the field of
  MACS~J0717.5+3745.}
\label{fig:histozMACS0717}
\end{figure}

\begin{figure}[h!] %fig C3
\centering 
\includegraphics[angle=0,width=6cm]{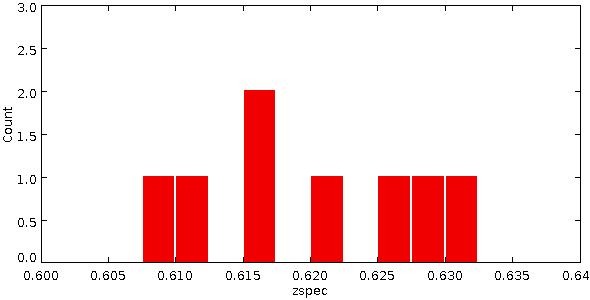}
%\mbox{\psfig{figure=Zw1332_GTC_NED_fin_histoz.ps,width=7cm}} %203.58
\caption{Galaxy spectroscopic redshift histogram in the field of
  ZwCl~1332.8+5043.}
\label{fig:histozZw1332}
\end{figure}

\begin{figure}[h!] %fig C4
\centering 
\includegraphics[angle=0,width=6cm]{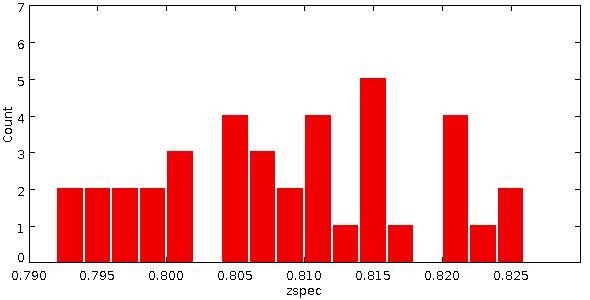}
%\mbox{\psfig{figure=RX1716_histoz.ps,width=7cm}} %259.2
\caption{Galaxy spectroscopic redshift histogram in the field of
  RX~J1716.4+6708.}
\label{fig:histozRX1716}
\end{figure}

\begin{figure}[h!] %fig C5
\centering 
\includegraphics[angle=0,width=6cm]{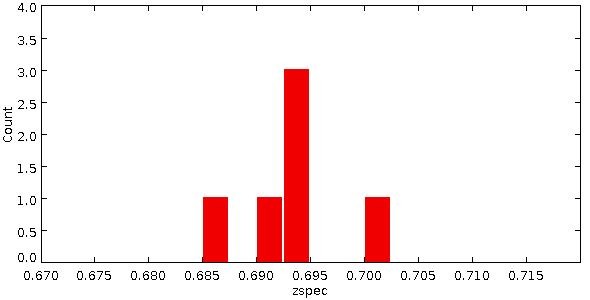}
%\mbox{\psfig{figure=NEP200_GTC_NED_fin_histoz.ps,width=7cm}} %269.33
\caption{Galaxy spectroscopic redshift histogram in the field of
  NEP~200.}
\label{fig:histozNEP200}
\end{figure}

\end{document}